\documentclass[hyper, a4paper, 11pt, oneside]{JHEP3} % 10pt is ignored!

\usepackage{amssymb,amsmath}
\usepackage{graphicx,epsfig}
\usepackage{epsfig,multicol,bbm}

\newcommand\fverb{\setbox\fverbbox=\hbox\bgroup\verb}
\newcommand\fverbdo{\egroup\medskip\noindent%
			\fbox{\unhbox\fverbbox}\ }
\newcommand\fverbit{\egroup\item[\fbox{\unhbox\fverbbox}]}
\newbox\fverbbox

\title{Particle Mixing and $CP$-Violation}

\author{Fayyazuddin\\
	National Centre for Physics and Physics department,
Quaid-i-Azam University, 45320 Islamabad, Pakistan\\
	E-mail: \email{fayyazuddins@gmail.com}}

\date{\today}

\abstract{In this review, the $X^{0}-\overline{X}^{0}$\ mixing ($%
X^{0}=B^{0},B_{s}^{0},K^{0}$) and its implication for CP\ violation in the
standard model are discussed. Both direct and mixing induced CP\ violation
for $K^{0}(\overline{K}^{0})$, $B^{0}(\overline{B}^{0})$\ and $B_{s}^{0}(%
\overline{B}_{s}^{0})$\ are reviewed.}


\begin{document}

\section{Introduction}

Symmetries have played an important role in particle physics. In quantum
mechanics a symmetry is associated with a group of transformations under
which a Lagrangian remains invariant. Symmetries limit the possible terms in
a Lagrangian and are associated with conservation laws. Here we will be
concerned with the role of discrete symmetries: Space Reflection (Parity) $P$%
: $\vec{x}\rightarrow -\vec{x}$, Time Reversal $T$: $t\rightarrow -t$ and
Charge Conjugation $C$: $particle\rightarrow antiparticle$.

Quantum Electrodynamics (QED) and Quantum Chromodynamics (QCD) respect all
these symmetries. Also, all Lorentz invariant local quantum field theories
are $CPT$ invariant. However, in weak interactions $C$ and $P$ are maximally
violated.

First indication of parity violation was revealed in the decay of a particle
with spin parity $J^{P}=0^{-},$ called $K$-meson into two modes $%
K^{0}\rightarrow \pi ^{+}\pi ^{-}$ (parity violating), and $K^{0}\rightarrow
\pi ^{+}\pi ^{-}$ $\pi ^{0}$(parity conserving).

Lee and Yang in 1956, suggested that there is no experimental evidence for
parity conservation in weak interaction. They suggested number of
experiments to test the validity of space reflection invariance in weak
decays. One way to test this is to measure the helicity of outgoing muon in
the decay:
\begin{equation*}
\pi ^{+}\rightarrow \mu ^{+}+\nu _{\mu }
\end{equation*}%
The helicity of muon comes out to be negative, showing that parity
conservation does not hold in this decay. In the rest frame of the pion,
since $\mu ^{+}$ comes out with negative helicity, the neutrino must also
come out with negative helicity because of the spin conservation. Thus
confirming the fact that neutrino is left handed.
\begin{equation*}
\pi ^{+}\rightarrow \mu ^{+}(-)+\nu _{\mu }
\end{equation*}%
Under charge conjugation,%
\begin{equation*}
\pi ^{+}\overset{C}{\rightarrow }\pi ^{-}\qquad \mu ^{+}\overset{C}{%
\rightarrow }\mu ^{-}\qquad \nu _{\mu }\overset{C}{\rightarrow }\bar{\nu}%
_{\mu }
\end{equation*}%
Helicity $\mathcal{H}=\frac{\vec{\sigma}\cdot \vec{p}}{\left\vert \vec{p}%
\right\vert }$ under $C$ and $P$ transforms as,%
\begin{equation*}
\mathcal{H}\overset{C}{\rightarrow }\mathcal{H},\qquad \mathcal{H}\overset{P}%
{\rightarrow }-\mathcal{H}
\end{equation*}%
Invariance under $C$ gives,%
\begin{equation*}
\Gamma _{\pi ^{+}\rightarrow \mu ^{+}(-)\nu _{\mu }}=\Gamma _{\pi
^{-}\rightarrow \mu ^{-}(-)\bar{\nu}_{\mu }}
\end{equation*}%
Experimentally,%
\begin{equation*}
\Gamma _{\pi ^{+}\rightarrow \mu ^{+}(-)\nu _{\mu }}>>\Gamma _{\pi
^{-}\rightarrow \mu ^{-}(-)\bar{\nu}_{\mu }}
\end{equation*}%
showing that $C$ is also violated in weak interactions. However, under $CP$,%
\begin{equation*}
\Gamma _{\pi ^{+}\rightarrow \mu ^{+}(-)\nu _{\mu }}\overset{CP}{\rightarrow
}\text{ \ }\Gamma _{\pi ^{-}\rightarrow \mu ^{-}(+)\bar{\nu}_{\mu }}
\end{equation*}%
which is seen experimentally. Thus, $CP$ conservation holds for this decay.

The $CP$ violaton in weak interaction is not universal, does not embrace all
weak processes unlike $C$ and $P$ violation. The $C$ and $P$ violation is
incorporated in the basic structure of theory by assigning the left-handed
and the right-handed fermions to doublet and singlet representations of the
elecroweak group $SU_{L}(2)\times U_{Y}(1)$%
\begin{eqnarray*}
\psi _{q} &=&\left(
\begin{array}{c}
u_{i} \\
d_{i}^{\prime }%
\end{array}%
\right) _{L};Y=1/3 \\
u_{iR} &:&Y=4/3 \\
d_{iR} &:&Y=-2/3 \\
\psi _{l} &=&\left(
\begin{array}{c}
\nu _{e^{-}} \\
e_{i}^{-}%
\end{array}%
\right) ;Y=-1 \\
e_{iR}^{-} &:&Y=2
\end{eqnarray*}%
Here $i$ is the generation index. The weak eigenstates $d^{\prime }$, $%
s^{\prime }$ and $b^{\prime }$ are different from the mass eigenstates $d$, $%
s$ and $b$. They are related to each other by a unitarity transformation,%
\begin{equation}
\left(
\begin{array}{c}
d^{\prime } \\
s^{\prime } \\
b^{\prime }%
\end{array}%
\right) =V\left(
\begin{array}{c}
d \\
s \\
b%
\end{array}%
\right)  \label{05}
\end{equation}%
where $V$ is called the $CKM$ matrix.%
\begin{equation*}
V=\left(
\begin{array}{ccc}
V_{ud} & V_{us} & V_{ub} \\
V_{cd} & V_{cs} & V_{cb} \\
V_{td} & V_{ts} & V_{tb}%
\end{array}%
\right)
\end{equation*}%
\begin{equation}
\simeq \left(
\begin{array}{ccc}
1-\frac{1}{2}\lambda ^{2} & \lambda & A\lambda ^{3}\left( \rho -i\eta \right)
\\
-\lambda & 1-\frac{1}{2}\lambda ^{2} & A\lambda ^{2} \\
A\lambda ^{3}\left( 1-\rho -i\eta \right) & -A\lambda ^{2} & 1%
\end{array}%
\right) +O\left( \lambda ^{4}\right) ,\text{ \ \ }\lambda =0.22  \label{06}
\end{equation}%
With three generations of quarks, there is one independent weak phase as
reflected with non zero value of $\eta $. The unitarity of $V$, $\left[
\text{Fig.1}\right] $ $VV^{\dagger }=1$ gives%
\begin{equation}
V_{ud}^{\ast }V_{ub}+V_{cb}^{\ast }V_{cd}+V_{td}^{\ast }V_{tb}=0  \label{07}
\end{equation}%
The second line in equation (\ref{06}) expresses $V$ in terms of Wolfenstien
parametrization. Thus,
\begin{eqnarray*}
V_{cb} &=&A\lambda ^{2} \\
V_{ub} &=&\left\vert V_{ub}\right\vert e^{-i\gamma } \\
V_{td} &=&\left\vert V_{td}\right\vert e^{-i\beta }
\end{eqnarray*}%
where,
\begin{equation*}
\tan {\gamma }=\frac{\eta }{\rho }=\frac{\bar{\eta}}{\bar{\rho}},\quad \tan {%
\beta }=\frac{\bar{\eta}}{1-\bar{\rho}},\quad \bar{\rho}=\rho (1-\frac{%
\lambda ^{2}}{2}),\quad \bar{\eta}=\eta (1-\frac{\lambda ^{2}}{2}).
\end{equation*}%
The weak angles $\beta $ and $\gamma $ play a leading role in $CP$
violation. However these weak angles are in $V_{ub}$\ and $V_{td}$, which
connect the first generation with the third generation. Hence the role of $%
\beta $ and $\gamma $\ in $K$ and $D$ decays is perepheral as both $K$ and $%
D $ are bound states of the first and second generation quarks.

The current in the standard model: $\bar{\Psi}_{i}\gamma ^{\mu }(1-\gamma
^{5})\Psi _{j}$, under $CP$ and time reversal transforms as
\begin{eqnarray*}
&&\bar{\Psi}_{i}\gamma ^{\mu }(1-\gamma ^{5})\Psi _{j}\overset{CP}{%
\rightarrow }-\eta \left( \mu \right) \bar{\Psi}_{j}\gamma ^{\mu }(1-\gamma
^{5})\Psi _{i} \\
&&\text{ \ \ \ \ \ \ \ \ \ \ \ \ \ \ \ \ \ \ \ \ \ \ \ \ }\overset{T}{%
\rightarrow }\eta \left( \mu \right) \bar{\Psi}_{i}\gamma ^{\mu }(1-\gamma
^{5})\Psi _{j}
\end{eqnarray*}%
where,
\begin{equation*}
\eta \left( \mu \right) =%
\begin{cases}
+1, & \text{if $\mu $=0} \\
-1, & \text{if $\mu $=1,2,3}%
\end{cases}%
\end{equation*}%
The Lagrangian,%
\begin{eqnarray*}
\mathcal{L}_{W} &=&\bar{\Psi}_{i}\gamma ^{\mu }(1-\gamma ^{5})\Psi
_{j}W_{\mu }^{+}+h.c. \\
&&\overset{CP}{\rightarrow }\bar{\Psi}_{j}\gamma ^{\mu }(1-\gamma ^{5})\Psi
_{i}W_{\mu }^{-}+h.c. \\
&&\overset{T}{\rightarrow }\bar{\Psi}_{i}\gamma ^{\mu }(1-\gamma ^{5})\Psi
_{j}W_{\mu }^{-}+h.c.
\end{eqnarray*}%
where%
\begin{equation*}
W_{\mu }^{\pm }\overset{CP}{\rightarrow }-\eta \left( \mu \right) W_{\mu
}^{\mp }\text{, }W_{\mu }^{\pm }\overset{T}{\rightarrow }\eta \left( \mu
\right) W_{\mu }^{\pm }
\end{equation*}%
Note that under $CP,$ $\bar{\Psi}_{i}\gamma ^{\mu }(1-\gamma ^{5})\Psi _{j}$%
\ goes over to its Hermitian conjugate. The flavor changing part of the
current is the charged current and contains the CKM\ matrix. The flavor non
changing part of the current is the neutral current, $CP$ violation is not
possible in weak processes involving neutral currents. Similarly in the
process involving the lepton current $\bar{\Psi}_{i}^{l}\gamma ^{\mu
}(1-\gamma ^{5})\Psi _{j}^{l}$, $CP$ is conserved.

\begin{figure}[tbp]
\begin{center}
\includegraphics[scale=0.5]{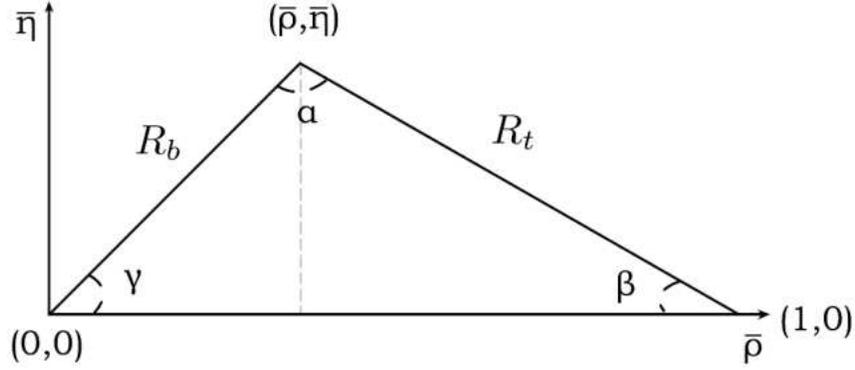}
\end{center}
\caption{Unitarity triangle $\protect\alpha +\protect\beta +\protect\gamma =%
\protect\pi $, $R_{b}=\protect\sqrt{\overline{\protect\rho }^{2}+\overline{%
\protect\eta }^{2}}$, $R_{t}=\protect\sqrt{(1-\overline{\protect\rho })^{2}+%
\overline{\protect\eta }^{2}}.$ }
\end{figure}

The following comments are in order. The quarks are basic constituent of
hadrons. Each quark has a definite charge, definite mass and definite
flavor. In a weak process, hadronic flavor changes. Hence \ the weak eigen
states can be a mixture of mass eigenstates. With three generations of
quarks, the weak eigenstates are related to the mass eigenstates by CKM
matrix. The CKM matrix also takes care of the experimental fact, the
suppression of weak processes from one generation to other.

The mismatch between the weak and mass eigenstates, involve the weak phase
in the CKM matrix. This can be a source of $CP$-violation in flavor changing
weak processes in the standard model.

In the standard model, the lepton number and the baryon number are conserved
$\tau _{e}>4.6\times 10^{26}yr$, $\tau _{p}>10^{31}yr$. However for leptons,
there is another conservation law: the lepton number for each generation is
separately conserved i.e. not only $\Delta L=0$, but $\Delta L_{e}=0$, $%
\Delta L_{\mu }=0$, $\Delta L_{\tau }=0$. For purely leptonic process, the
limit on flavor changing processes $\mu \rightarrow e\gamma $, $\tau
\rightarrow \mu (e)\gamma $ is $\Gamma (\mu \rightarrow e\gamma )/\Gamma
(\mu \rightarrow all)<1.2\times 10^{-11}$. Hence for charged leptons, there
is a stringent limit on flavor changing processes $\Delta L_{e}\neq 0$, $%
\Delta L_{\mu }\neq 0$ and $\Delta L_{\tau }\neq 0$. Thus for charged
leptons, there is no difference between weak and mass eigenstates. However
for neutrios separate conservation of lepton number is not required, as a
consequence neutrino mixing analogous to CKM quark mixing is possible. This
results in neutrino oscillations. Neutrinos are stable particles, the
neutrino mixing plays no role in $CP$ violation. The $CP$ violation in
lepton sector will violate lepton number. Hence $CP$ violation in lepton
sector falls in the same catagory as $CP$ violation required for
baryogenesis along with $\Delta B\neq 0$.

Under charge conjugation, a particle is transformed to its antiparticle.
Since in the standard model, the electric charge, baryon and lepton number
are conserved; hence for $CP$ eigenstates formed from the states with $Q=0$,
$B=0$, $L=0$, $\Delta B=0=\Delta L$. However for neutrinos, and neutral
baryons, $CP$ eigenstates would have $\Delta L=2$\ and $\Delta B=2$, hence
not allowed in the standard model. Thus only $CP$-eigenstates for neutral
mesons viz $X^{0}\equiv (K^{0}$, $D^{0}$, $B^{0}$, $B_{s}^{0}\dot{)}$ are
allowed. Now under $CP$:
\begin{equation*}
\left\vert X^{0}\right\rangle \overset{CP}{\rightarrow }\eta
_{X}^{CP}\left\vert \overline{X}^{0}\right\rangle
\end{equation*}%
where $\eta _{X}^{CP}$\ is the $CP$-phase. We select $\eta _{X}^{CP}=-1$,\
with this convention, the $CP$-eigenstates are%
\begin{equation*}
\left\vert X_{1,2}^{0}\right\rangle =\frac{1}{\sqrt{2}}\left[ \left\vert
X^{0}\right\rangle \mp \left\vert \overline{X}^{0}\right\rangle \right]
\text{, }CP=\pm 1
\end{equation*}%
In the weak interaction, both hypercharge and isospin are violated, so only $%
CP$-eigenstates can be mass eigenstates when weak interaction Hamiltonion is
included in the Hamiltonian. When weak interaction is switched off; the mass
eigenstates are $\left\vert X^{0}\right\rangle $\ and $\left\vert \overline{X%
}^{0}\right\rangle $, with same mass and same lifetime, a consequence of
CPT\ theorem. To summarize

\begin{itemize}
\item For $X^{0}-\overline{X}^{0}$\ complex ($X^{0}=K^{0},B^{0},B_{s}^{0}$);
the mass matrix is not diagonal in $\left\vert X^{0}\right\rangle $\ and $%
\left\vert \overline{X}^{0}\right\rangle $ basis.

\item However, assuming $CP$ conservation, the $CP$ eigenstates $\left\vert
X_{1}^{0}\right\rangle $\ and $\left\vert X_{2}^{0}\right\rangle $\ can be
mass eigenstates and hence mass matrix is diagonal in this basis.

\item The two sets of states are related to each other by superposition
principle of quantum mechanics.

\item This gives rise to quantum mechanical interference so that even if we
start with a state $\left\vert X^{0}\right\rangle $, the time evolution of
this state can generate $\left\vert \overline{X}^{0}\right\rangle $. This is
a source of mixing induced $CP$ violation.

\item However, both in $K^{0}-\overline{K}^{0}$\ and $B^{0}-\overline{B}^{0}$%
\ complex, the mass eigenstates $\left\vert K_{S}^{0}\right\rangle $, $%
\left\vert K_{L}^{0}\right\rangle $\ and $\left\vert B_{S}^{0}\right\rangle $%
, $\left\vert B_{L}^{0}\right\rangle $\ are not $CP$ eigenstates.

\item In the case of $K^{0}-\overline{K}^{0}$\ complex,%
\begin{equation*}
\left\vert K_{S}^{0}\right\rangle =\left\vert K_{1}^{0}\right\rangle
+\epsilon \left\vert K_{2}^{0}\right\rangle
\end{equation*}%
\begin{equation*}
\left\vert K_{L}^{0}\right\rangle =\left\vert K_{2}^{0}\right\rangle
+\epsilon \left\vert K_{1}^{0}\right\rangle
\end{equation*}%
there is a small admixture of wrong $CP$ state characterized by small
parameter $\epsilon $, which gives rise to the $CP$ violating decay $%
K_{L}^{0}\rightarrow \pi ^{+}\pi ^{-}$. This was the first $CP$ violating
decay observed experimentally.

\item For $D^{0}-\overline{D}^{0}$\ complex, there is no mismatch between $%
CP $ eigenstates $\left\vert D_{1}^{0}\right\rangle ,\left\vert
D_{2}^{0}\right\rangle $\ and mass eigen states; $\left\vert
D^{0}\right\rangle ,\left\vert \overline{D}^{0}\right\rangle $ are bound
states of $1^{st}$ and $2^{nd}$ generations of quarks-antiquarks.

\item For $B^{0}-\overline{B}^{0}$\ complex, the mismatch between mass
eigenstates and $CP$ eigenstates $\left\vert B_{1}^{0}\right\rangle $ and $%
\left\vert B_{2}^{0}\right\rangle $ is given by the phase factor $e^{2i\phi
_{M}}$\ where the pase factor $\phi _{M}=-\beta $\ in the standard model
viz. one of the angle in the CKM matrix.%
\begin{eqnarray*}
\left\vert B_{L}^{0}\right\rangle &=&\frac{1}{\sqrt{2}}\left[ \left\vert
B^{0}\right\rangle -e^{2i\phi _{M}}\left\vert \overline{B}^{0}\right\rangle %
\right] \\
\left\vert B_{H}^{0}\right\rangle &=&\frac{1}{\sqrt{2}}\left[ \left\vert
B^{0}\right\rangle +e^{2i\phi _{M}}\left\vert \overline{B}^{0}\right\rangle %
\right]
\end{eqnarray*}

\item For $B_{s}^{0}-\overline{B}_{s}^{0}$, there is no mismatch between $CP$
eigenstates $\left\vert B_{1s}^{0}\right\rangle $ and $\left\vert
B_{2s}^{0}\right\rangle $\ and the mass eigenstates. With three generations
of quarks, no extra phase is available to generate mismatch between mass and
$CP$ eigensates for $B_{s}^{0}-\overline{B}_{s}^{0}$ complex.

\item The quantum mechanical interference gives rise to non zero mass
difference $\Delta m_{K}$, $\Delta m_{B}$\ and $\Delta m_{B_{s}}$\ between
mass eigenstates. The mixing induced $CP$ violation involves these mass
differences.
\end{itemize}

\section{$CPT$ and $CP$ invariance}

It is instructive to discuss the restrictions imposed by $CPT$ invariance. $%
CPT$ invariance implies,
\begin{eqnarray}
_{\text{out}}\left\langle f\left\vert \mathcal{L}\right\vert X\right\rangle
&=&_{\text{out}}\left\langle f\left\vert \left( CPT\right) ^{-1}\mathcal{L}%
CPT\right\vert X\right\rangle  \notag \\
&=&\eta _{T}^{X\ast }\eta _{T}^{f}\,\,_{\text{in}}\left\langle \tilde{f}%
\left\vert \left( CP\right) ^{\dagger }\mathcal{L}^{\dagger }\left(
CP\right) ^{-1\dagger }\right\vert \widetilde{X}\right\rangle ^{\ast }
\notag \\
&=&\eta _{T}^{X\ast }\eta _{T}^{f}\left\langle \widetilde{X}\left\vert
\left( CP\right) ^{-1}\mathcal{L}\left( CP\right) \right\vert \widetilde{f}%
\right\rangle _{\text{in}}  \notag
\end{eqnarray}%
where \symbol{126}means momentum and spin of the final state are reversed;
\symbol{126}may be droped. Further, we may choose the $CP$ phase such that%
\begin{equation}
CP\left\vert X\right\rangle =-\left\vert \bar{X}\right\rangle  \label{5}
\end{equation}%
\begin{equation}
CP\left\vert f\right\rangle =\eta _{f}^{CP}\left\vert \bar{f}\right\rangle
\label{6}
\end{equation}%
Thus we have%
\begin{equation}
_{\text{out}}\left\langle f\left\vert \mathcal{L}\right\vert X\right\rangle
=\eta _{f}\left\langle \overline{X}\left\vert \mathcal{L}\right\vert \bar{f}%
\right\rangle _{\text{in}}
\end{equation}%
where%
\begin{equation}
\eta _{f}=-\eta _{CP}^{f}\eta _{T}^{\ast }\eta _{T}^{X\ast }
\end{equation}%
Hence on using%
\begin{equation}
\left\vert f\right\rangle _{\text{in}}=S_{f}\left\vert f\right\rangle _{%
\text{out}}=\exp (2i\delta _{f})\left\vert f\right\rangle _{\text{in}}
\label{7}
\end{equation}%
we get%
\begin{eqnarray}
_{\text{out}}\left\langle f\left\vert \mathcal{L}\right\vert X\right\rangle
&=&\eta _{f}\,e^{2i\delta _{f}}\left\langle \overline{X}\left\vert \mathcal{L%
}\right\vert \bar{f}\right\rangle _{\text{out}}  \label{8} \\
&=&\eta _{f}\,e^{2i\delta _{f}}{}_{\text{out}}\left\langle \bar{f}\left\vert
\mathcal{L}\right\vert \overline{X}\right\rangle ^{\ast }
\end{eqnarray}%
Hence finally we have%
\begin{equation}
\bar{A}_{\bar{f}}=\eta _{f}e^{2i\delta _{f}}A_{f}^{\ast }  \label{8b}
\end{equation}%
If $CP$-invariance holds, then,%
\begin{equation}
_{\text{out}}\left\langle f\left\vert \mathcal{L}\right\vert X\right\rangle
=_{\text{ out}}\left\langle \bar{f}\left\vert \mathcal{L}\right\vert \bar{X}%
\right\rangle \Rightarrow \bar{A}_{\bar{f}}=A_{f}.
\end{equation}%
Thus, the necessary condition for $CP$-violation is that the decay amplitude
$A$ should be complex. In view of our discussion, under $CP$ an operator $%
O\left( \vec{x},t\right) $ is replaced by,%
\begin{equation}
O\left( \vec{x},t\right) \rightarrow O^{\dagger }\left( -\vec{x},t\right)
\label{9}
\end{equation}%
and under time reversal%
\begin{equation}
O\left( \vec{x},t\right) \rightarrow O\left( \vec{x},-t\right)
\end{equation}%
The effective Lagrangian has the structure ($\mathcal{L}^{\dagger }=\mathcal{%
L}$),%
\begin{eqnarray}
\mathcal{L} &=&aO+a^{\ast }O^{\dagger }  \label{10} \\
&&\overset{CP}{\rightarrow }aO^{\dagger }+a^{\ast }O  \notag \\
&&\overset{T}{\rightarrow }a^{\ast }O+aO^{\dagger }  \notag \\
&&\overset{CPT}{\rightarrow }a^{\ast }O^{\dagger }+aO=\mathcal{L}  \notag
\end{eqnarray}%
Hence, $CP$-violation requires $a^{\ast }\neq a$. We now discuss the
implication of $CPT$ constraint with respect to $CP$ violation of weak
decays. The weak amplitude is complex; it contains the final state strong
phase $\delta _{f}$ and in addition it may also contain a weak phase $\phi $%
. Taking out both these phases,%
\begin{equation*}
A_{f}=e^{i\phi }e^{i\delta _{f}}\left\vert A_{f}\right\vert
\end{equation*}%
$CPT$ Eq \eqref{8b} gives,%
\begin{equation*}
\bar{A}_{\bar{f}}=\eta _{f}e^{2i\delta _{f}}A_{f}^{\ast }=\eta _{f}e^{-i\phi
}e^{i\delta _{f}}\left\vert A_{f}\right\vert
\end{equation*}%
For direct $CP$ violation, at least two amplitudes with different weak
phases are required%
\begin{equation}
A_{f}=A_{1f}+A_{2f}
\end{equation}%
$CPT$ gives%
\begin{eqnarray*}
\bar{A}_{\bar{f}} &=&e^{2i\delta _{1f}}A_{1f}^{\ast }+e^{2i\delta
_{2f}}A_{2f}^{\ast } \\
A_{if} &=&e^{i\phi _{i}}e^{i\delta _{if}}\left\vert A_{if}\right\vert
\end{eqnarray*}%
where ($\delta _{1f},\delta _{2f}$), ($\phi _{1},\phi _{2}$) are strong
final state phases and the weak phases respectively. Thus the direct $CP$
violation is given by%
\begin{eqnarray}
A_{CP} &=&\frac{\overline{\Gamma }(\overline{X}\rightarrow \overline{f}%
)-\Gamma (X\rightarrow f)}{\overline{\Gamma }(\overline{X}\rightarrow
\overline{f})+\Gamma (X\rightarrow f)}  \notag \\
&=&\frac{2\left\vert A_{1f}\right\vert \left\vert A_{2f}\right\vert \sin
\phi \sin \delta _{f}}{\left\vert A_{1f}\right\vert ^{2}+\left\vert
A_{2f}\right\vert ^{2}+2\left\vert A_{1f}\right\vert \left\vert
A_{2f}\right\vert \cos \phi \cos \delta _{f}}
\end{eqnarray}%
where $\delta _{f}=\delta _{2f}-\delta _{1f}$, $\phi =\phi _{1}-\phi _{2}$%
.Hence the necessary condition for non zero direct $CP$ violaton is $\delta
_{f}\neq 0$ and $\phi \neq 0$.\ The weak phase may be a consequence of phase
in CKM matrix.

\section{Particle Mixing}

In $\left\vert X^{0}\right\rangle -\left\vert \bar{X}^{0}\right\rangle $
basis,%
\begin{eqnarray*}
|\psi (t)\rangle  &=&a(t)|X^{0}\rangle +\bar{a}(t)|\bar{X}^{0}\rangle  \\
\frac{i}{dt}|\psi (t)\rangle  &=&M|\psi (t)\rangle
\end{eqnarray*}%
The mass matrix $M$ is not diagonal and is given by,%
\begin{equation}
M=m-\frac{i}{2}\Gamma =\left(
\begin{array}{cc}
m_{11}-\frac{i}{2}\Gamma _{11} & m_{12}-\frac{i}{2}\Gamma _{12} \\
m_{21}-\frac{i}{2}\Gamma _{21} & m_{22}-\frac{i}{2}\Gamma _{22}%
\end{array}%
\right)   \label{18}
\end{equation}%
Hermiticity of matrices $m_{\alpha \alpha ^{\prime }}$ and $\Gamma _{\alpha
\alpha ^{\prime }}$ gives ($\alpha =\alpha ^{\prime }=1,2$),
\begin{eqnarray}
\left( m\right) _{\alpha \alpha ^{\prime }} &=&\left( m^{\dagger }\right)
_{\alpha \alpha ^{\prime }}=\left( m^{\ast }\right) _{\alpha ^{\prime
}\alpha },\qquad \Gamma _{\alpha \alpha ^{\prime }}=\Gamma _{\alpha ^{\prime
}\alpha }^{\ast }  \notag \\
m_{21} &=&m_{12\,}^{\ast }\qquad \Gamma _{21}=\Gamma _{12}^{\ast }
\end{eqnarray}%
$CPT$ invariance gives,%
\begin{equation}
\left\langle X^{0}\left\vert M\right\vert X^{0}\right\rangle =\left\langle
\bar{X}^{0}\left\vert M\right\vert \bar{X}^{0}\right\rangle   \notag
\end{equation}%
\begin{equation*}
m_{11}=m_{22},\qquad \Gamma _{11}=\Gamma _{22}
\end{equation*}%
\begin{equation}
\left\langle \bar{X}^{0}\left\vert M\right\vert X^{0}\right\rangle
=\left\langle \bar{X}^{0}\left\vert M\right\vert X^{0}\right\rangle \text{:
identity}
\end{equation}%
Diagonalization of mass matrix $M$ in eq. (\ref{18}) gives,%
\begin{eqnarray}
m_{11}-\frac{i}{2}\Gamma _{11}-pq &=&m_{1}-\frac{i}{2}\Gamma _{1}  \notag \\
m_{11}-\frac{i}{2}\Gamma _{11}+pq &=&m_{2}-\frac{i}{2}\Gamma _{2}  \label{21}
\end{eqnarray}%
where,
\begin{equation}
p^{2}=m_{12}-\frac{i}{2}\Gamma _{12},\qquad q^{2}=m_{12}^{\ast }-\frac{i}{2}%
\Gamma _{12}^{\ast }
\end{equation}%
The eigenstates are given by,%
\begin{equation}
|X_{1,2}\rangle =\frac{1}{\sqrt{\left\vert p\right\vert ^{2}+\left\vert
q\right\vert ^{2}}}\left[ p|X^{0}\rangle \mp q|\bar{X}^{0}\rangle \right]
\label{23}
\end{equation}%
From Eq (\ref{21}), taking the real and imaginary parts, we have%
\begin{eqnarray*}
m_{1} &=&m_{11}-\text{Re}pq \\
m_{2} &=&m_{11}-\text{Re}pq \\
\Gamma _{1} &=&\Gamma _{11}+2\text{Im}pq \\
\Gamma _{2} &=&\Gamma _{11}-2\text{Im}pq
\end{eqnarray*}%
Thus finally we have%
\begin{eqnarray}
\Delta m &=&m_{2}-m_{1}=2\text{Re}pq  \notag \\
m &=&\frac{m_{1}+m_{2}}{2}=m_{11}  \notag \\
\Delta \Gamma  &=&\Gamma _{2}-\Gamma _{1}=-4\text{Im}pq  \notag \\
\Gamma  &=&\frac{\Gamma _{1}+\Gamma _{2}}{2}=\Gamma _{11}  \label{A}
\end{eqnarray}%
Let us define%
\begin{eqnarray}
q/p &=&\frac{1-\epsilon }{1+\epsilon }  \notag \\
&=&\sqrt{\frac{m_{12}^{\ast }-\frac{i}{2}\Gamma _{12}^{\ast }}{m_{12}^{{}}-%
\frac{i}{2}\Gamma _{12}^{{}}}}  \label{B}
\end{eqnarray}%
It follows that CP-violation is determined by the parameter%
\begin{equation}
\epsilon =\frac{p-q}{p+q}  \label{C}
\end{equation}%
Now $\left\vert X_{1}\right\rangle $ and $\left\vert X_{2}\right\rangle $
are mass eigenstates. They form a complete set (in units $\hbar =c=1$),%
\begin{eqnarray}
\left\vert \psi \left( t\right) \right\rangle  &=&a\left( t\right)
\left\vert X_{1}\right\rangle +b\left( t\right) \left\vert
X_{2}\right\rangle   \notag \\
i\frac{d\left\vert \psi \left( t\right) \right\rangle }{dt} &=&\left(
\begin{array}{cc}
m_{1}-\frac{i}{2}\Gamma _{1} & 0 \\
0 & m_{2}-\frac{i}{2}\Gamma _{2}%
\end{array}%
\right) \left\vert \psi \left( t\right) \right\rangle .  \label{2.1}
\end{eqnarray}%
The solution is,%
\begin{eqnarray*}
a\left( t\right)  &=&a\left( 0\right) \exp \left( -im_{1}t-\frac{1}{2}\Gamma
_{1}t\right)  \\
b\left( t\right)  &=&b\left( 0\right) \exp \left( -im_{2}t-\frac{1}{2}\Gamma
_{2}t\right)
\end{eqnarray*}%
Suppose we start with the state $\left\vert X^{0}\right\rangle $, i.e.,%
\begin{equation*}
\left\vert \psi \left( 0\right) \right\rangle =\left\vert X^{0}\right\rangle
\end{equation*}%
After time $t$%
\begin{eqnarray}
\left\vert \psi \left( t\right) \right\rangle  &=&\frac{\sqrt{\left\vert
p\right\vert ^{2}+\left\vert q\right\vert ^{2}}}{2p}\left[ \exp \left(
-im_{1}t-\frac{1}{2}\Gamma _{1}t\right) \left\vert X_{1}\right\rangle+ \exp \left( -im_{2}t-\frac{1}{2}\Gamma _{2}t\right) \left\vert
X_{2}\right\rangle \right]   \notag \\
&=&\frac{1}{2}\left\{ \left[ \exp \left( -im_{1}t-\frac{1}{2}\Gamma
_{1}t\right) \right. +\exp \left( -im_{2}t-\frac{1}{2}\Gamma _{2}t\right) \right]
\left\vert X^{0}\right\rangle   \notag \\
&&-\frac{q}{p}\left[ \exp \left( -im_{1}t-\frac{1}{2}\Gamma _{1}t\right)
\left. -\exp \left( -im_{2}t-\frac{1}{2}\Gamma _{2}t\right) \right]
\left\vert \bar{X}^{0}\right\rangle \right\}   \label{iii}
\end{eqnarray}%
Eq (\ref{iii}) clearly shows the particle mixing. Similarly if we start with
$\left\vert \bar{X}^{0}\right\rangle $ we get after time t%
\begin{eqnarray}
\left\vert \psi \left( t\right) \right\rangle  &=&\frac{1}{2}\left\{ \frac{p%
}{q}\left[ \exp \left( -im_{1}t-\frac{1}{2}\Gamma _{1}t\right) \right. -\exp \left( -im_{2}t-\frac{1}{2}\Gamma _{2}t\right) \right]
\left\vert X^{0}\right\rangle   \notag \\
&&-\left[ \exp \left( -im_{1}t-\frac{1}{2}\Gamma _{1}t\right)  \left. +\exp \left( -im_{2}t-\frac{1}{2}\Gamma _{2}t\right) \right]
\left\vert \bar{X}^{0}\right\rangle \right\}   \label{iv}
\end{eqnarray}%
From Eqs. (\ref{iii}) and (\ref{iv}), we can determine $X^{0}$\ and $\bar{X}%
^{0}$\ mixing. It is clear that if we start with $X^{0},$\ then at time t,
the probability of finding the particles $X^{0}$\ or $\bar{X}^{0}$ is given
by [using Eq (\ref{iii})]%
\begin{equation}
\left\vert \left\langle X^{0}|\psi (t)\right\rangle \right\vert ^{2}=\frac{1%
}{4}\left[ e^{-\Gamma _{1}t}+e^{-\Gamma _{2}t}+2e^{-\Gamma t}\cos \Delta mt%
\right]   \label{v}
\end{equation}%
\begin{equation}
\left\vert \left\langle \bar{X}^{0}|\psi (t)\right\rangle \right\vert ^{2}=%
\frac{1}{4}\left\vert \frac{1-\epsilon }{1+\epsilon }\right\vert ^{2}\left[
e^{-\Gamma _{1}t}+e^{-\Gamma _{2}t}-2e^{-\Gamma t}\cos \Delta mt\right]
\label{vi}
\end{equation}%
We define the mixing parameter $r$\ as
\begin{equation}
r=\frac{\int_{0}^{T}\left\vert \left\langle \overline{X}^{0}|\psi
(t)\right\rangle \right\vert ^{2}dt}{\int_{0}^{T}\left\vert \left\langle
X^{0}|\psi (t)\right\rangle \right\vert ^{2}dt}
\end{equation}%
where $T$ is a sufficiently long time. In the limit $T\rightarrow \infty $,
using Eqs (\ref{v}) and (\ref{vi}), we get%
\begin{equation}
r=\left\vert \frac{1-\epsilon }{1+\epsilon }\right\vert ^{2}\frac{x^{2}+y^{2}%
}{2+x^{2}-y^{2}}
\end{equation}%
where $x=\frac{\Delta m}{\Gamma }$\ and $y=\frac{\Delta \Gamma }{2\Gamma }$.
If we start with $\bar{X}^{0},$ we can use Eq (\ref{iv})\ then%
\begin{equation}
\overline{r}=\frac{\int_{0}^{T}\left\vert \left\langle X^{0}|\psi
(t)\right\rangle \right\vert ^{2}dt}{\int_{0}^{T}\left\vert \left\langle
\overline{X}^{0}|\psi (t)\right\rangle \right\vert ^{2}dt}\underset{%
T\rightarrow \infty }{\longrightarrow }\left\vert \frac{1+\epsilon }{%
1-\epsilon }\right\vert ^{2}\frac{x^{2}+y^{2}}{2+x^{2}-y^{2}}
\end{equation}%
When $CP$-violation effects are neglected, then
\begin{equation}
r=\overline{r}=\frac{x^{2}+y^{2}}{2+x^{2}-y^{2}}
\end{equation}%
The asymmetry parameter a%
\begin{equation}
a=\frac{\overline{r}-r}{\overline{r}+r}=\frac{4\text{Re}\epsilon }{%
1+\left\vert \epsilon \right\vert ^{2}}
\end{equation}%
is a measure of $CP$-violation. We define another parameter \ which is also
a measure of particle mixing. Let $\chi $\ be the probability $%
X^{0}\rightarrow \bar{X}^{0}$, then
\begin{equation}
\chi =\int_{0}^{T}\left\vert \left\langle \overline{X}^{0}|\psi
(t)\right\rangle \right\vert ^{2}dt  \notag
\end{equation}%
\begin{equation}
1-\chi =\int_{0}^{T}\left\vert \left\langle X^{0}|\psi (t)\right\rangle
\right\vert ^{2}dt  \notag
\end{equation}%
Thus%
\begin{equation}
r=\frac{\chi }{1-\chi },\chi =\frac{r}{1+r}
\end{equation}%
Similarly, we get%
\begin{equation}
\overline{r}=\frac{\overline{\chi }}{1-\overline{\chi }},\overline{\chi }=%
\frac{\overline{r}}{1+\overline{r}}
\end{equation}%
We note from definitions, $x=\Delta m/\Gamma $, $y=\Delta \Gamma /\Gamma $
\begin{eqnarray*}
0 &\leq &x^{2}\leq \infty  \\
0 &\leq &y^{2}\leq 1
\end{eqnarray*}%
obviously%
\begin{equation*}
0\leq r\leq 1
\end{equation*}

\bigskip

\section{$K^{0}-\bar{K}^{0}$ Complex and $CP$--Violation in $K$-Decay}

Since,
\begin{equation*}
CP\left( \pi ^{+}\,\pi ^{-}\right) =\left( -1\right) ^{l}\left( -1\right)
^{l}=1
\end{equation*}%
therefore, it is clear that,
\begin{equation*}
K_{1}^{0}\longrightarrow \pi ^{+}\,\pi ^{-}
\end{equation*}%
is allowed by $CP$ conservation.

However, experimentally it was found that long lived $K_{2}^{0}$ also decay
to $\pi ^{+}\,\pi ^{-}$ but with very small probability. Small $CP$ non
conservation can be taken into account by defining,
\begin{eqnarray}
\left\vert K_{S}\right\rangle &=&\left\vert K_{1}^{0}\right\rangle
+\varepsilon \left\vert K_{2}^{0}\right\rangle  \notag \\
\left\vert K_{L}\right\rangle &=&\left\vert K_{2}^{0}\right\rangle
+\varepsilon \left\vert K_{1}^{0}\right\rangle  \label{2.11}
\end{eqnarray}%
where $\varepsilon $ is a small number. Thus $CP$ non conservation manifests
itself by the ratio:
\begin{eqnarray}
\eta _{+-} &=&\frac{A\left( K_{L}\rightarrow \pi ^{+}\,\pi ^{-}\right) }{%
A\left( K_{S}\rightarrow \pi ^{+}\,\pi ^{-}\right) }=\varepsilon
\label{2.12} \\
\left\vert \eta _{+-}\right\vert &\simeq &\left( 2.286\pm 0.017\right)
\times 10^{-3}  \notag
\end{eqnarray}%
Now $CP$ non conservation implies,
\begin{equation}
m_{12}\neq m_{12}^{\ast },\qquad \Gamma _{12}\neq \Gamma _{12}^{\ast }.
\label{2.13}
\end{equation}%
Since $CP$ violation is a small effect, therefore,
\begin{equation}
\text{Im}m_{12}\ll \text{Re}m_{12}\qquad \text{Im}\Gamma _{12}\ll \text{Re}%
\Gamma _{12}.  \label{2.14}
\end{equation}%
Further, if $CP$- violation arises from mass matrix, then,
\begin{equation}
\Gamma _{12}=\Gamma _{12}^{\ast }.  \label{2.15}
\end{equation}%
For small $\epsilon $\
\begin{eqnarray}
q^{2}/p^{2} &=&1-4\epsilon  \notag \\
4\epsilon &=&\frac{p^{2}-q^{2}}{p^{2}}
\end{eqnarray}%
Now
\begin{equation*}
p^{2},q^{2}\approx \left( \text{Re}m_{12}-\frac{i}{2}\Gamma _{12}\right) %
\left[ 1\pm \frac{i\text{Im}m_{12}}{\text{Re}m_{12}-\frac{i}{2}\Gamma _{12}}%
\right]
\end{equation*}%
Hence
\begin{eqnarray}
\epsilon &=&\frac{i\text{Im}m_{12}}{2\left( \text{Re}m_{12}-\frac{i}{2}%
\Gamma _{12}\right) }  \notag \\
&=&\frac{i\text{Im}m_{12}}{\left( m_{2}-m_{1}\right) -i\left( \Gamma
_{2}-\Gamma _{1}\right) }
\end{eqnarray}%
Then from Eq. (\ref{A}) up to first order, we get,
\begin{eqnarray}
\Delta m &=&m_{2}-m_{1}\rightarrow m_{K_{L}}-m_{K_{S}}  \notag \\
&=&2\text{Re}m_{12}  \notag \\
\Delta \Gamma &=&\Gamma _{2}-\Gamma _{1}=\Gamma _{L}-\Gamma _{S}=2\Gamma
_{12}  \label{2.18}
\end{eqnarray}%
Now,
\begin{eqnarray}
\Delta m &=&m_{L}-m_{S}  \notag \\
\Delta \Gamma &=&\Gamma _{L}-\Gamma _{S}  \notag \\
\Gamma _{S} &=&\frac{\hbar }{\tau _{S}}=7.367\times 10^{-12}\text{ MeV}%
,\,\,\,  \notag \\
&&\left. \tau _{S}=\left( 0.8935\pm 0.0008\right) \times 10^{-10}\text{ s}%
\right.  \notag \\
\Gamma _{L} &=&\frac{\hbar }{\tau _{L}}=1.273\times 10^{-14}\text{ MeV},\,\,
\notag \\
&&\left. \,\tau _{L}=\left( 5.17\pm 0.04\right) \times 10^{-8}\text{ s}%
\right.  \notag \\
\Delta \Gamma &\simeq &-\Gamma _{S}  \notag \\
m_{L} &=&m+\frac{1}{2}\Delta m  \notag \\
m_{S} &=&m-\frac{1}{2}\Delta m  \label{2.19}
\end{eqnarray}%
Hence from Eq. (\ref{iii}),%
\begin{equation}
\left\vert \psi \left( t\right) \right\rangle \approx \frac{1}{2}e^{\frac{-i%
}{2}mt}\left\{
\begin{array}{c}
\left[ e^{\frac{-1}{2}\Gamma _{S}t}e^{\frac{i}{2}\Delta mt}+e^{-\frac{i}{2}%
\Delta mt}\right] \left\vert K^{0}\right\rangle \\
-\left[ e^{\frac{-1}{2}\Gamma _{S}t}e^{\frac{i}{2}\Delta mt}-e^{-\frac{i}{2}%
\Delta mt}\right] \left\vert \bar{K}^{0}\right\rangle%
\end{array}%
\right\}  \label{2.20}
\end{equation}%
Therefore, probability of finding $\bar{K}^{0}$ at time $t$ (recall that we
started with $K^{0}$),%
\begin{eqnarray}
P\left( K^{0}\rightarrow \bar{K}^{0},t\right) &=&\left\vert \left\langle
\bar{K}^{0}\left\vert {}\right. \psi \left( t\right) \right\rangle
\right\vert ^{2}  \notag \\
&=&\frac{1}{4}\left( 1+e^{-\Gamma _{S}t}-2e^{-\frac{1}{2}\Gamma _{S}t}\cos
\left( \Delta m\right) t\right)  \notag \\
&=&\frac{1}{4}\left( 1+e^{-t/\tau _{S}}-2e^{-\frac{1}{2}t/\tau _{S}}\cos
\left( \Delta m\right) t\right)  \notag \\
&&  \label{2.21}
\end{eqnarray}%
If kaons were stable $(\tau _{S}\rightarrow \infty )$, then,%
\begin{equation}
P\left( K^{0}\rightarrow \bar{K}^{0},t\right) =\frac{1}{2}\left[ 1-\cos
\left( \Delta m\right) t\right]  \label{2.22}
\end{equation}%
which shows that a state produced as pure $Y=1$ state at $t=0$ continuously
oscillates between $Y=1$ and $Y=-1$ state with frequency $\omega =\frac{%
\Delta m}{\hbar }$ and period of oscillation,%
\begin{equation}
\tau =\frac{2\pi }{\left( \Delta m/\hbar \right) }.  \label{2.23}
\end{equation}%
Kaons, however, decay and their oscillations are damped.

By measuring the period of oscillation, $\Delta m$ can be determined.%
\begin{eqnarray}
\Delta m &=&m_{L}-m_{S}=\left( 3.483\pm 0.006\right) \times 10^{-12}\text{
MeV.}  \label{2.24} \\
&=&047\Gamma _{s}
\end{eqnarray}%
Such a small number is measured as a consequence of superposition principle
of quantum mechanics.

Coming back to $CP$-violation,
\begin{eqnarray}
\varepsilon &=&\frac{i\text{Im}m_{12}}{\Delta m-i\Delta \Gamma /2}\qquad
\varepsilon =\left\vert \varepsilon \right\vert e^{i\phi _{\varepsilon }}
\label{2.25} \\
\tan \phi _{\varepsilon } &=&-2\Delta m/\Delta \Gamma =\Delta m/\Gamma
_{S}-\Gamma _{L}  \notag \\
&\approx &\frac{2\times 0.474\Gamma _{S}}{0.998\Gamma _{S}}  \notag \\
&\Rightarrow &\phi _{\varepsilon }=(43.51\pm 0.05)^{\circ }  \label{2.26} \\
\left\vert \epsilon \right\vert &=&(2.228\pm 0.011)\times 10^{-3}
\label{2.226}
\end{eqnarray}%
So far we have considered $CP$-violation due to mixing in the mass matrix.
It is important to detect the $CP$-violation in the decay amplitude if any.
This is done by looking for a difference between $CP$-violation for the
final $\pi ^{0}\pi ^{0}$ and $\pi ^{+}\pi ^{-}$ states. Now due to Bose
statistics, the two pions can be either in $I=0$ or $I=2$ states. Using
Clebsch-Gordon (CG) coefficients,
\begin{eqnarray}
A\left( K^{0}\rightarrow \pi ^{+}\pi ^{-}\right) &=&\frac{1}{\sqrt{3}}\left[
\sqrt{2}A_{0}e^{i\delta _{0}}+A_{2}e^{i\delta _{2}}\right]  \notag \\
A\left( K^{0}\rightarrow \pi ^{0}\pi ^{0}\right) &=&\frac{1}{\sqrt{3}}\left[
A_{0}e^{i\delta _{0}}-\sqrt{2}A_{2}e^{i\delta _{2}}\right]  \label{2.27}
\end{eqnarray}%
Now $CPT$-invariance gives,
\begin{eqnarray}
A\left( \bar{K}^{0}\rightarrow \pi ^{+}\pi ^{-}\right) &=&\frac{1}{\sqrt{3}}%
\left[ \sqrt{2}A_{0}^{\ast }e^{i\delta _{0}}+A_{2}^{\ast }e^{i\delta _{2}}%
\right]  \notag \\
A\left( \bar{K}^{0}\rightarrow \pi ^{0}\pi ^{0}\right) &=&\frac{1}{\sqrt{3}}%
\left[ A_{0}^{\ast }e^{i\delta _{0}}-\sqrt{2}A_{2}^{\ast }e^{i\delta _{2}}%
\right]  \label{2.28}
\end{eqnarray}%
The dominant decay amplitude is $A_{0}$ due to $\Delta I=1/2$ rule, $%
\left\vert A_{2}/A_{0}\right\vert \simeq 1/22$. Using the Wu--Yang phase
convention, we can take $A_{0}$ to be real. Neglecting terms of order $%
\varepsilon \text{Re}\frac{A_{2}}{A_{0}}$ and $\varepsilon \text{Im}\frac{%
A_{2}}{A_{0}}$, we get,
\begin{eqnarray}
\eta _{+-} &\equiv &\left\vert \eta _{+-}\right\vert e^{i\phi _{+-}}\simeq
\varepsilon +\varepsilon ^{\prime }  \notag \\
\eta _{00} &\equiv &\left\vert \eta _{00}\right\vert e^{i\phi _{00}}\simeq
\varepsilon -2\varepsilon ^{\prime }  \label{2.29}
\end{eqnarray}%
where,
\begin{equation}
\varepsilon ^{\prime }=\frac{i}{\sqrt{2}}e^{i\left( \delta _{2}-\delta
_{0}\right) }\text{Im}\frac{A_{2}}{A_{0}}  \label{2.30}
\end{equation}%
Clearly $\varepsilon ^{\prime }$ measures the $CP$-violation in the decay
amplitude, since $CP$-invariance implies $A_{2}$ to be real.

After $35$ years of experiments at Fermilab and CERN, results have converged
on a definitive non-zero result for $\varepsilon ^{\prime }$,
\begin{eqnarray}
R &=&\left\vert \frac{\eta _{00}}{\eta _{+-}}\right\vert ^{2}=\left\vert
\frac{\varepsilon -2\varepsilon ^{\prime }}{\varepsilon +\varepsilon
^{\prime }}\right\vert ^{2},\qquad \varepsilon ^{\prime }\ll \varepsilon
\notag \\
&\simeq &\left\vert 1-\frac{3\varepsilon ^{\prime }}{\varepsilon }%
\right\vert ^{2}\simeq 1-6\text{Re}\left( \varepsilon ^{\prime }/\varepsilon
\right)  \notag \\
\text{Re}\left( \varepsilon ^{\prime }/\varepsilon \right) &=&\frac{1-R}{6}
\label{2.31} \\
&=&\left( 1.65\pm 0.26\right) \times 10^{-3}.  \label{2.32}
\end{eqnarray}%
\begin{equation}
\sin (\phi _{00}-\phi _{+-})=3\text{Re}(\epsilon ^{\prime }/\epsilon )\tan
(\phi _{\epsilon }-\phi _{\epsilon ^{\prime }})  \label{2.33}
\end{equation}%
This is an evidence that although $\varepsilon ^{\prime }$ is a very small,
but $CP$-violation does occur in the decay amplitude. Further we note from
Eq. (\ref{2.30}),
\begin{equation*}
\phi _{\varepsilon ^{\prime }}=\delta _{2}-\delta _{0}+\frac{\pi }{2}\approx
42.3\pm 1.5^{0}
\end{equation*}%
where numerical value is based on an analysis of $\pi \pi $ scattering. The
experimental values of $CP$-violation parameters are as follows%
\begin{eqnarray}
\left\vert \eta _{+-}\right\vert &=&(2.233\pm 0.010)\times 10^{-3}  \notag \\
\left\vert \eta _{00}\right\vert &=&(2.222\pm 0.010)\times 10^{-3}
\end{eqnarray}%
\begin{eqnarray}
\phi _{+-} &=&(43.4\pm 0.007)^{\circ }  \notag \\
\phi _{00} &=&(43.7\pm 0.008)^{\circ }
\end{eqnarray}%
$CPT$ invariance predicts (Cf Eq. \ref{2.33})%
\begin{equation}
\phi _{00}-\phi _{+-}\approx 3\text{Re}(\epsilon ^{\prime }/\epsilon )\text{
}(\phi _{\epsilon }-\phi _{\epsilon ^{\prime }})\approx 0
\end{equation}%
We now discuss the $CP$-asymmetry in leptonic decays of kaon.
\begin{eqnarray*}
\frac{\Delta S}{\Delta Q} &=&1 \\
K^{+} &\rightarrow &\pi ^{0}+l^{+}+\nu _{l} \\
K^{0} &\rightarrow &\pi ^{-}+l^{+}+\nu _{l}=f \\
\overline{K}^{0} &\rightarrow &\pi ^{+}+l^{-}+\overline{\nu }_{l}=f^{\ast }%
\text{ $CPT$} \\
\frac{\Delta S}{\Delta Q} &=&-1 \\
K^{0} &\rightarrow &\pi ^{+}+l^{-}+\overline{\nu }_{l}=g^{\ast } \\
\overline{K}^{0} &\rightarrow &\pi ^{-}+l^{+}+\nu _{l}=g\text{ $CPT$}
\end{eqnarray*}%
\begin{eqnarray*}
A(K_{L}^{0} &\rightarrow &\pi ^{-}+l^{+}+\nu _{l})=\frac{1}{\sqrt{2}}%
[(1+\epsilon )f+(1-\epsilon )g] \\
A(K_{L}^{0} &\rightarrow &\pi ^{+}+l^{-}+\overline{\nu }_{l})=\frac{1}{\sqrt{%
2}}[(1+\epsilon )g^{\ast }+(1-\epsilon )f\ast ]
\end{eqnarray*}%
The $CP$-asymmetry parameter $\delta _{l}:$%
\begin{eqnarray}
\delta _{l} &=&\frac{\Gamma (K_{L}^{0}\rightarrow \pi ^{-}l^{+}\nu
_{l})-\Gamma (K_{L}^{0}\rightarrow \pi ^{+}l^{-}\overline{\nu }_{l})}{\Gamma
(K_{L}^{0}\rightarrow \pi ^{-}l^{+}\nu _{l})+\Gamma (K_{L}^{0}\rightarrow
\pi ^{+}l^{-}\overline{\nu }_{l})}  \notag \\
&=&\frac{2\text{Re}\epsilon \lbrack \left\vert f\right\vert ^{2}-\left\vert
g\right\vert ^{2}]}{\left\vert f\right\vert ^{2}+\left\vert g\right\vert
^{2}+(fg^{\ast }+f^{\ast }g)+O(\epsilon ^{2})}
\end{eqnarray}%
In the standard model $\frac{\Delta S}{\Delta Q}=-1$ transitions are not
allowed, thus $g=0$. Hence
\begin{equation*}
\delta _{l}\approx 2\text{Re}\epsilon =(3.32\pm 0.06)10^{-3}[\text{Expt.
value}]
\end{equation*}%
From Eq. (\ref{2.25}), we get
\begin{equation*}
2\text{Re}\epsilon =2\left\vert \epsilon \right\vert \cos \phi _{\epsilon }
\end{equation*}%
which gives on using expermintal values for $\left\vert \epsilon \right\vert
$ and $\phi _{\epsilon }$%
\begin{equation*}
2\text{Re}\epsilon =(3.23\pm 0.02\times 10^{-3})
\end{equation*}%
in agreement with the expermimental value for $\delta _{l}$. The
experimental value of $\delta _{l}$\ shows the internal consistancy of the
standard model and the $CPT$ invariance.

Finally we discuss $CP$-asymmetries for $K\rightarrow 3\pi $ decays. The
decays
\begin{eqnarray*}
K^{+} &\rightarrow &\pi ^{+}\pi ^{0}\pi ^{0}\text{, }\pi ^{+}\pi ^{+}\pi ^{-}
\\
K^{0} &\rightarrow &\pi ^{+}\pi ^{-}\pi ^{0}\text{, }\pi ^{0}\pi ^{0}\pi ^{0}
\end{eqnarray*}%
are partiy conserving decays i.e. the parity of the final state is $-1$. Now
the C-partiy of $\pi ^{0}$ and ($\pi ^{+}\pi ^{-})_{l^{\prime }}$ are given
by
\begin{equation*}
C(\pi ^{0})=1,\text{ }C(\pi ^{+}\pi ^{-})=(-1)^{l^{\prime }}
\end{equation*}%
and G-parity of pion is $-1.$ Thus
\begin{eqnarray*}
CP|\pi ^{0}\pi ^{0}\pi ^{0} &>&=-|\pi ^{0}\pi ^{0}\pi ^{0}> \\
CP|\pi ^{+}\pi ^{-}\pi ^{0} &>&=(-1)^{l^{\prime }+1}|\pi ^{+}\pi ^{-}\pi
^{0}>
\end{eqnarray*}%
Hence $CP$-conservation implies
\begin{eqnarray*}
K_{2}^{0} &\rightarrow &\text{ }\pi ^{0}\pi ^{0}\pi ^{0}\text{ allowed.} \\
K_{1}^{0} &\rightarrow &\text{ }\pi ^{0}\pi ^{0}\pi ^{0}\text{ is forbidden.}
\\
K_{1}^{0} &\rightarrow &\pi ^{+}\pi ^{-}\pi ^{0}\text{ allowed if }l^{\prime
}\text{ is odd.} \\
K_{2}^{0} &\rightarrow &\pi ^{+}\pi ^{-}\pi ^{0}\text{ allowed if }l^{\prime
}\text{ is even.}
\end{eqnarray*}%
Now G-partiy of three pions $\pi ^{+}\pi ^{-}\pi ^{0}:$%
\begin{eqnarray*}
G &=&C(-1)^{I}=(-1)^{l^{\prime }+I}=-1 \\
\text{Hence }l^{\prime } &=&\text{even},\text{ }I(\text{odd});\text{ }I=1,3
\\
l^{\prime } &=&\text{odd},\text{ }I(\text{even});\text{ }I=0,2
\end{eqnarray*}%
Only $l^{\prime }=0$ decays are favored as the decays for $l^{\prime }>0$
are highly suppressed due to centrifugal barrier. Hence $K_{1}^{0}%
\rightarrow \pi ^{+}\pi ^{-}\pi ^{0}$ is highly suppressed. Thus we have to
take into account $I=1,3$ amplitudes viz $a_{1}$ and $a_{3}$. $I=3$
contribution is expected to be suppressed as it requires $\Delta I=\frac{5}{2%
}$ transition.

Hence $CP$-asymmetries of $K^{0}\rightarrow 3\pi $ decays are given by
\begin{eqnarray*}
\eta _{000} &=&\frac{A(K_{S}\rightarrow \text{ }\pi ^{0}\pi ^{0}\pi ^{0})}{%
A(K_{L}\rightarrow \text{ }\pi ^{0}\pi ^{0}\pi ^{0})}=\frac{%
(a_{1}-a_{1}^{\ast })+\epsilon (a_{1}+a_{1}^{\ast })}{(a_{1}+a_{1}^{\ast
})+\epsilon (a_{1}-a_{1}^{\ast })}=\frac{[i\text{Im}a_{1}+\epsilon \text{Re}%
a_{1}]}{\text{Re}a_{1}+i\epsilon \text{Im}a_{1}} \\
&\approx &\epsilon +i\frac{\text{Im}a_{1}}{\text{Re}a_{1}} \\
\eta _{+-0} &=&\frac{A(K_{S}\rightarrow \pi ^{+}\pi ^{-}\pi ^{0})}{%
A(K_{L}\rightarrow \pi ^{+}\pi ^{-}\pi ^{0})}\approx \epsilon +i\frac{\text{%
Im}a_{1}}{\text{Re}a_{1}}=\eta _{000}
\end{eqnarray*}

\section{$B^{0}-\bar{B}^{0}$ Complex}

For $B_{q}^{0}$ (q=d or s) we show below that both $m_{12}$ and $\Gamma _{12}
$ have the same phase. Thus,
\begin{eqnarray}
m_{12} &=&\left\vert m_{12}\right\vert e^{-2i\phi _{M}}  \notag \\
\Gamma _{12} &=&\left\vert \Gamma _{12}\right\vert e^{-2i\phi _{M}}
\label{3.1} \\
\left\vert \Gamma _{12}\right\vert  &\ll &\left\vert m_{12}\right\vert
\notag \\
p^{2} &=&e^{-2i\phi _{M}}\left[ \left\vert m_{12}\right\vert -i\left\vert
\Gamma _{12}\right\vert \right] \simeq \left\vert m_{12}\right\vert
e^{-2i\phi _{M}}  \notag \\
q^{2} &=&e^{+2i\phi _{M}}\left[ \left\vert m_{12}\right\vert -i\left\vert
\Gamma _{12}\right\vert \right] \simeq \left\vert m_{12}\right\vert
e^{2i\phi _{M}}  \label{3.2} \\
q/p &=&e^{-2i\phi _{M}}  \label{3.0} \\
2pq &=&2\left\vert m_{12}\right\vert =(m_{2}-m_{1})-\frac{i}{2}\left( \Gamma
_{2}-\Gamma _{1}\right)   \notag \\
\Rightarrow \Delta m_{B} &=&\left( m_{2}-m_{1}\right) =2\left\vert
m_{12}\right\vert   \label{3.3} \\
\Delta \Gamma  &=&\Gamma _{2}-\Gamma _{1}=0  \notag
\end{eqnarray}

For $B_{d}:\phi _{M}=-\beta $

For $B_{s}:\phi _{M}=0$

The above equations follow from the fact that,
\begin{equation*}
m_{12}-i\Gamma _{12}=\langle \bar{B}_{q}^{0}\left\vert H_{eff}^{\Delta
B=2}\right\vert B_{q}^{0}\rangle
\end{equation*}%
$H_{eff}^{\Delta B=2}$ induces particle-antiparticle transition. For $%
B_{q}^{0}\rightarrow \bar{B}_{q}^{0},$ $H_{eff}^{\Delta B=2}$ arises from
the box diagram as shown in Fig. 2, where the dominant contribution comes
out from the $t-$quark. Thus,
\begin{equation*}
m_{12}\varpropto (V_{tb})^{2}(V_{tq}^{\ast })^{2}m_{t}^{2}
\end{equation*}

\begin{figure}[tbp]
\begin{center}
\includegraphics[scale=0.5]{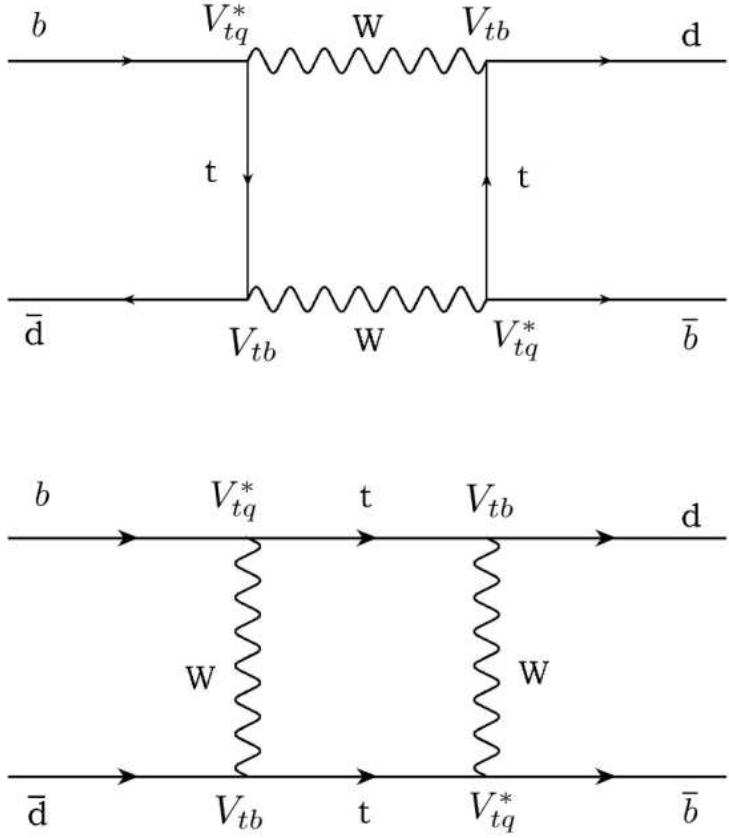}
\end{center}
\caption{Box Diagrams}
\end{figure}

Now,
\begin{equation*}
\Gamma _{12}\varpropto \sum_{f}\langle \bar{B}^{0}\left\vert
H_{W}\right\vert f\rangle \langle f\left\vert H_{W}\right\vert B^{0}\rangle
\end{equation*}%
where the sum is over all the final states which contribute to both $%
B_{q}^{0}$ and $\bar{B}_{q}^{0}$ decays. Thus,
\begin{equation*}
\Gamma _{12}\varpropto \left( V_{cb}V_{cq}^{\ast }+V_{ub}V_{uq}^{\ast
}\right) ^{2}m_{b}^{2}\propto (V_{tb})^{2}(V_{tq}^{\ast })^{2}m_{b}^{2}
\end{equation*}%
Hence we have the result that,
\begin{equation*}
\frac{\left\vert \Gamma _{12}\right\vert }{\left\vert m_{12}\right\vert }%
\sim \frac{m_{b}^{2}}{m_{t}^{2}}
\end{equation*}%
Now $B_{d}^{0}\rightarrow \bar{B}_{d}^{0}$ transition:
\begin{equation*}
\left( V_{tb}\right) ^{2}\left( V_{td}^{\ast }\right) ^{2}=A^{2}\lambda ^{6}
\left[ \left( 1+\rho \right) ^{2}+\eta ^{2}\right] e^{2i\beta }
\end{equation*}%
Hence,
\begin{equation*}
m_{12}=\left\vert m_{12}\right\vert e^{2i\beta },\qquad \Gamma
_{12}=\left\vert \Gamma _{12}\right\vert e^{2i\beta },\qquad \phi _{M}=-\beta
\end{equation*}%
On the other hand, $B_{s}^{0}\rightarrow \bar{B}_{s}^{0}$ transition:%
\begin{equation}
\left( V_{tb}\right) ^{2}\left( V_{ts}^{\ast }\right) ^{2}=\left\vert
V_{ts}\right\vert ^{2}\approx A^{2}\lambda ^{4}  \label{3.26}
\end{equation}%
\begin{equation}
m_{12}=\left\vert m_{12}\right\vert ,\qquad \Gamma _{12}=\left\vert \Gamma
_{12}\right\vert  \label{3.27}
\end{equation}%
\begin{equation}
\phi _{M}=0  \label{3.28}
\end{equation}%
Also we have,%
\begin{equation}
\frac{\Delta m_{B_{s}}}{\Delta m_{B_{d}}}=\frac{\left\vert m_{12}\right\vert
_{s}}{\left\vert m_{12}\right\vert _{d}}=\frac{1}{\lambda ^{2}\left[ \left(
1-\overline{\rho }\right) ^{2}+\overline{\eta }^{2}\right] }\xi \approx 34\xi
\label{3.29}
\end{equation}%
where $\xi $ is $SU(3)$ breaking parameter. The numerical value is obtained
using the experimental values $\lambda =0.225$, $\overline{\rho }=0.132$, $%
\overline{\eta }=0.341$.

Hence it follows from Eqs. (\ref{23}), (\ref{3.2}) and (\ref{3.0}) the mass
eigenstates $B_{L}^{0}$ and $B_{H}^{0}$ can be written as:
\begin{eqnarray}
\left\vert B_{L}^{0}\right\rangle  &=&\frac{1}{\sqrt{2}}\left[ \left\vert
B^{0}\right\rangle -e^{2i\phi _{M}}\left\vert \bar{B}^{0}\right\rangle %
\right] \quad CP=+1,\phi _{M}\rightarrow 0  \notag \\
&&  \label{3.4} \\
\left\vert B_{H}^{0}\right\rangle  &=&\frac{1}{\sqrt{2}}\left[ \left\vert
B^{0}\right\rangle +e^{2i\phi _{M}}\left\vert \bar{B}^{0}\right\rangle %
\right] \quad CP=-1,\phi _{M}\rightarrow 0  \notag \\
&&  \label{3.5}
\end{eqnarray}%
In this case, $CP$ violation occurs due to phase factor $e^{2i\phi _{M}}$ in
the mass matrix.

Now one gets (from Eq. (\ref{iii})), using Eqs.(\ref{3.3}), (\ref{3.4}) and (%
\ref{3.5}),
\begin{eqnarray}
\left\vert B^{0}\left( t\right) \right\rangle &=&e^{-imt}e^{-\frac{1}{2}%
\Gamma t}\left\{ \cos \left( \frac{\Delta m}{2}t\right) \left\vert
B^{0}\right\rangle \right.  \notag \\
&&\left. -ie^{+2i\phi _{M}}\sin \left( \frac{\Delta m}{2}t\right) \left\vert
\bar{B}^{0}\right\rangle \right\}  \label{3.8}
\end{eqnarray}%
Similarly we get,
\begin{eqnarray}
\left\vert \bar{B}^{0}\left( t\right) \right\rangle &=&-e^{-imt}e^{-\frac{1}{%
2}\Gamma t}\left\{ \cos \left( \frac{\Delta m}{2}t\right) \left\vert \bar{B}%
^{0}\right\rangle \right.  \notag \\
&&\left. -ie^{-2i\phi _{M}}\sin \left( \frac{\Delta m}{2}t\right) \left\vert
B^{0}\right\rangle \right\}  \label{3.9}
\end{eqnarray}%
Suppose we start with $B^{0}$ viz $|B^{0}\left( 0\right) \rangle
=|B^{0}\rangle ,$ the probabilities of finding $\bar{B}^{0}$ and $B^{0}$ at
time $t$ is given by,
\begin{eqnarray*}
P\left( B^{0}\rightarrow \bar{B}^{0},t\right) &=&\left\vert \langle \bar{B}%
^{0}|B^{0}\left( t\right) \rangle \right\vert ^{2} \\
&=&\frac{1}{2}e^{-\Gamma t}\left( 1-\cos (\Delta m\right) t) \\
P\left( B^{0}\rightarrow B^{0},t\right) &=&\left\vert \langle
B^{0}|B^{0}\left( t\right) \rangle \right\vert ^{2} \\
&=&\frac{1}{2}e^{-\Gamma t}\left( 1+\cos (\Delta m\right) t)
\end{eqnarray*}%
These are equations of a damped harmonic oscillator, the angular frequency
of which is,
\begin{equation*}
\omega =\frac{\Delta m}{\hslash }
\end{equation*}%
Now the mixing parameter,
\begin{eqnarray}
r &=&\frac{\int_{0}^{T}\left\vert \langle \bar{B}^{0}|B^{0}\left( t\right)
\rangle \right\vert ^{2}dt}{\int_{0}^{T}\left\vert \langle B^{0}|B^{0}\left(
t\right) \rangle \right\vert ^{2}dt}=\frac{\chi }{1-\chi }  \notag \\
&\xrightarrow{T\rightarrow \infty}&\frac{\left( \Delta m/\Gamma \right) ^{2}%
}{2+\left( \Delta m/\Gamma \right) ^{2}}=\frac{x^{2}}{2+x^{2}}
\end{eqnarray}%
Experimentally, for $B_{d}^{0}$ and $B_{s}^{0}$,
\begin{subequations}
\begin{eqnarray}
\Delta m_{B_{d}^{0}} &=&(0.507\pm 0.005)\times 10^{12}\hbar s^{-1}=(3.337\pm
0.033)\times 10^{-10}\text{MeV}  \notag \\
\tau _{B_{d}^{0}} &=&(1.525\pm 0.009)\times 10^{-12}s \\
\Delta m_{B_{s}^{0}} &=&(17.77\pm 0.12)\times 10^{12}\hbar s^{-1}=(117\pm
0.8)\times 10^{-10}\text{MeV}  \notag \\
\tau _{B_{s}^{0}} &=&(1.472\pm _{0.0026}^{0.0024})\times 10^{-12}s \\
x_{d} &=&\left( \frac{\Delta m_{B_{d}^{0}}}{\Gamma _{B_{d}^{0}}}\right)
=0.77\pm 0.008  \label{3.9b} \\
x_{s} &=&\left( \frac{\Delta m_{B_{s}^{0}}}{\Gamma _{B_{s}^{0}}}\right)
=26.2\pm 0.5
\end{eqnarray}%
Non zero values of $x_{d}$\ and $x_{s}$\ clearly show mixing between $B_{q},$
$B_{\overline{q}}$($q=s,$ $d$). The large value of the $x_{s}$\ compared to $%
x_{d}$\ is inconformity with Eq. (\ref{3.29}).

From Eq. (\ref{3.8}) and (\ref{3.9}), the decay amplitudes for,
\end{subequations}
\begin{eqnarray}
B^{0}\left( t\right) &\rightarrow &f\,\,\,\,\,\,\,\,\,\,\,\,\,A_{f}\left(
t\right) =\left\langle f\left\vert H_{w}\right\vert B^{0}\left( t\right)
\right\rangle  \notag \\
\bar{B}^{0}\left( t\right) &\rightarrow &\bar{f}\,\,\,\,\,\,\,\,\,\,\,\,\bar{%
A}_{\bar{f}}\left( t\right) =\left\langle \bar{f}\left\vert H_{w}\right\vert
\bar{B}^{0}\left( t\right) \right\rangle  \label{3.10}
\end{eqnarray}%
are given by,
\begin{eqnarray}
A_{f}\left( t\right) &=&e^{-imt}e^{-\frac{1}{2}\Gamma t}\left\{ \cos
\left( \frac{\Delta m}{2}t\right) A_{f} -ie^{+2i\phi _{M}}\sin \left( \frac{\Delta m}{2}%
t\right) \bar{A}_{f}\right\}  \label{3.11} \\
\bar{A}_{\bar{f}}\left( t\right) &=&e^{-imt}e^{-\frac{1}{2}\Gamma t}\left\{
\cos \left( \frac{\Delta m}{2}t\right) \bar{A}_{\bar{f}}-ie^{-2i\phi _{M}}\sin \left( \frac{\Delta m}{2}t\right) A_{\bar{f}%
}\right\} .  \label{3.12}
\end{eqnarray}%
From Eqs.(\ref{3.11}) and (\ref{3.12}), we get for the decay rates,
\begin{eqnarray}
\Gamma _{f}(t) &=&e^{-\Gamma t}\left[
\begin{array}{c}
\frac{1}{2}\left( \left\vert A_{f}\right\vert ^{2}+\left\vert \bar{A}%
_{f}\right\vert ^{2}\right) +\frac{1}{2}\left( \left\vert A_{f}\right\vert
^{2}-\left\vert \bar{A}_{f}\right\vert ^{2}\right) \cos \Delta mt \\
-\frac{i}{2}\left( 2i\text{Im}e^{2i\phi _{M}}A_{f}^{\ast }\bar{A}_{f}\right)
\sin \Delta mt%
\end{array}%
\right]  \notag \\
&&  \label{I} \\
\bar{\Gamma}_{\bar{f}}(t) &=&e^{-\Gamma t}\left[
\begin{array}{c}
\frac{1}{2}\left( \left\vert A_{\bar{f}}\right\vert ^{2}+\left\vert \bar{A}_{%
\bar{f}}\right\vert ^{2}\right) -\frac{1}{2}\left( \left\vert A_{\bar{f}%
}\right\vert ^{2}-\left\vert \bar{A}_{\bar{f}}\right\vert ^{2}\right) \cos
\Delta mt \\
+\frac{i}{2}\left( 2i\text{Im}e^{2i\phi _{M}}A_{\bar{f}}^{\ast }\bar{A}_{%
\bar{f}}\right) \sin \Delta mt%
\end{array}%
\right]  \notag \\
&&  \label{II}
\end{eqnarray}%
For $\Gamma _{\bar{f}}$ and $\bar{\Gamma}_{f}$ change $f\rightarrow \bar{f}$
and $\bar{f}\rightarrow f$ in $\Gamma _{f}$ and $\bar{\Gamma}_{\bar{f}}$
respectively. As a simple application of the above equations, consider the
semi-leptonic decays of $B^{0}$,
\begin{eqnarray*}
B^{0} &\rightarrow &l^{+}\nu X^{-}:f\text{ \ for example }X^{-}=D^{-} \\
\bar{B}^{0} &\rightarrow &l^{-}\bar{\nu}X^{+}:\bar{f}\text{ \ for example }%
X^{+}=D^{+}
\end{eqnarray*}%
In the standard model, $\bar{B}^{0}$ decay into $l^{+}\nu X^{-}$ and $B^{0}$
decay into $l^{-}\bar{\nu}X^{+}$ is forbidden. Thus,
\begin{eqnarray}
\bar{A}_{f} &=&0,\qquad A_{\bar{f}}=0  \notag \\
\Gamma _{f}(t) &=&e^{-\Gamma t}\frac{1}{2}\left\vert A_{f}\right\vert
^{2}\left( 1+\cos \Delta mt\right)  \notag \\
\Gamma _{\bar{f}}(t) &=&e^{-\Gamma t}\frac{1}{2}\left\vert \bar{A}_{\bar{f}%
}\right\vert ^{2}\left( 1-\cos \Delta mt\right) ,\because \left\vert \bar{A}%
_{\bar{f}}\right\vert =\left\vert A_{f}\right\vert
\end{eqnarray}%
Hence,
\begin{equation}
\delta =\frac{\int_{0}^{\infty }\Gamma _{\bar{f}}(t)dt}{\int_{0}^{\infty
}\Gamma _{f}(t)dt}=\frac{x_{d}^{2}}{2+x_{d}^{2}}=r_{d}
\end{equation}%
Non zero value of $\delta $ would indicate mixing. If, however, $\bar{A}%
_{f}\neq 0$ and $A_{\bar{f}}\neq 0$ due to some exotic mechanism, then $%
\delta \neq 0$ even without mixing. Thus
\begin{eqnarray}
\frac{\Gamma \left( \mu ^{-}X^{+}\right) }{\Gamma \left( \mu
^{+}X^{-}\right) +\Gamma \left( \mu ^{-}X^{+}\right) } &=&\frac{r_{d}}{%
1+r_{d}}=\chi _{d} \\
&=&0.172\pm 0.010\text{ (Expt value)}  \notag
\end{eqnarray}%
which gives,
\begin{equation*}
x_{d}=0.723\pm 0.032
\end{equation*}%
in agreement with $x_{d}$\ given in Eq (\ref{3.9b}).

\section{$CP$-Violation in $B$-Decays}

\subsection{Case I}

In this section, we discuss the $CP$-violation for $B\rightarrow f,\overline{%
f}$\ where%
\begin{equation*}
|\bar{f}\rangle =CP|f\rangle =|f\rangle
\end{equation*}%
For this case we get, from Eqs. (\ref{3.11}) and (\ref{3.12}),
\begin{eqnarray}
\mathcal{A}_{f}\left( t\right) &=&\frac{\Gamma _{f}\left( t\right) -\bar{%
\Gamma}_{f}\left( t\right) }{\Gamma _{f}\left( t\right) +\bar{\Gamma}%
_{f}\left( t\right) }=\cos \left( \Delta mt\right) \left( \left\vert
A_{f}\right\vert ^{2}-\left\vert \bar{A}_{f}\right\vert ^{2}\right)  \notag
\\
&&-i\sin \left( \Delta mt\right) \left( e^{2i\phi _{M}}A_{f}^{\ast }\bar{A}%
_{f}-e^{-2i\phi _{M}}A_{f}\bar{A}_{f}^{\ast }\right) /\left( \left\vert
A_{f}\right\vert ^{2}+\left\vert \bar{A}_{f}\right\vert ^{2}\right)  \notag
\\
&&  \label{3.13} \\
&=&\cos \left( \Delta mt\right) C_{f}+\sin \left( \Delta mt\right) S_{f}
\label{3.14}
\end{eqnarray}%
where,
\begin{equation}
C_{f}=\frac{1-\left\vert \bar{A}_{f}\right\vert ^{2}/\left\vert
A_{f}\right\vert ^{2}}{1+\left\vert \bar{A}_{f}\right\vert ^{2}/\left\vert
A_{f}\right\vert ^{2}}=\frac{1-\left\vert \lambda \right\vert ^{2}}{%
1+\left\vert \lambda \right\vert ^{2}}\qquad \lambda =\frac{\bar{A}_{f}}{%
A_{f}}  \label{3.14b}
\end{equation}%
This is the direct $CP$ violation and,
\begin{equation}
S_{f}=\frac{2\text{Im}\left( e^{2i\phi _{M}}\lambda \right) }{1+\left\vert
\lambda \right\vert ^{2}}  \label{93}
\end{equation}%
is the mixing induced $CP$-violation.

If the decay proceeds through a single diagram (for example tree graph), $%
\bar{A}_{f}/A_{f}$ is given by $CPT$,
\begin{equation}
\lambda =\frac{\bar{A}_{f}}{A_{f}}=\frac{e^{i\left( \phi +\delta _{f}\right)
}}{e^{i\left( -\phi +\delta _{f}\right) }}=e^{2i\phi }
\end{equation}%
where $\phi $ is the weak phase in the decay amplitude. Hence from Eq. (\ref%
{3.13}), we obtain,
\begin{equation}
\mathcal{A}_{f}(t)=\sin \left( \Delta mt\right) \sin \left( 2\phi _{M}+2\phi
\right)
\end{equation}

\begin{figure}[tbp]
\includegraphics[scale=0.5]{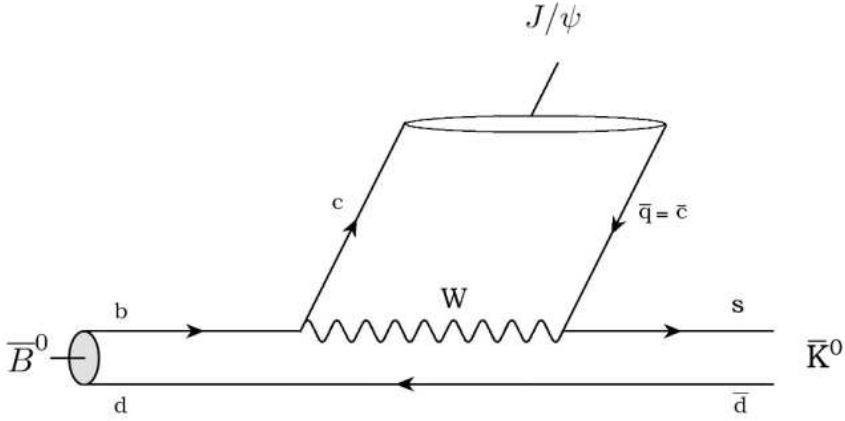}
\caption{Color suppressed tree diagram for $\overline{B}^{0}\rightarrow J/%
\protect\psi \,\overline{K}^{0}$}
\end{figure}

In particular for the decay (Fig 3),
\begin{equation*}
B^{0}\rightarrow J/\psi \,K_{s},\,\,\,\,\phi =0
\end{equation*}%
we obtain,
\begin{equation}
\mathcal{A}_{\psi K_{s}}\left( t\right) =\sin \left( 2\phi _{M}\right) \sin
\left( \Delta mt\right) =-\sin 2\beta \sin (\Delta mt)  \label{3.15}
\end{equation}%
and,
\begin{eqnarray}
\mathcal{A}_{\psi K_{s}} &=&\frac{\int_{0}^{\infty }\left[ \Gamma _{f}\left(
t\right) -\bar{\Gamma}_{f}\left( t\right) \right] dt}{\int_{0}^{\infty }%
\left[ \Gamma _{f}\left( t\right) +\bar{\Gamma}_{f}\left( t\right) \right] dt%
}  \notag \\
\mathcal{A}_{\psi K_{s}} &=&-\sin \left( 2\beta \right) \,\,\frac{\left(
\Delta m/\Gamma \right) }{1+\left( \Delta m/\Gamma \right) ^{2}}
\label{3.16} \\
\text{Experiment} &:&\left( \frac{\Delta m}{\Gamma }\right)
_{B_{d}^{0}}=0.776\pm 0.008  \label{3.17}
\end{eqnarray}%
$\mathcal{A}_{\psi K_{s}}$ has been experimentally measured. It gives,
\begin{equation*}
\sin 2\beta =0.678\pm 0.025
\end{equation*}%
Corresponding to the decay $B^{0}\rightarrow J/\psi \,K_{s}$, we have the
decay $B_{s}^{0}\rightarrow J/\psi \,\phi .$ Thus for this decay
\begin{eqnarray}
\mathcal{A}_{J/\psi \phi }^{(t)} &=&-\sin 2\beta _{s}\sin (\Delta m_{B_{s}}t)
\\
\mathcal{A}_{J/\psi \phi } &=&-\sin 2\beta _{s}\frac{(\Delta
m_{B_{s}}/\Gamma _{s})}{1+(\Delta m_{B_{s}}/\Gamma _{s})^{2}}
\end{eqnarray}%
In the standard model, $\beta _{s}=0,$ $\mathcal{A}_{J/\psi \phi }=0.$This
is an example of $CP$-violation in the mass matrix. Non zero value of $\mathcal{A}_{J/\psi \phi }$ will
be a signal for physics beyond standard model. We now discuss the direct $CP$%
-violation.

Direct $CP$-violation in $B$ decays involves the weak phase in the decay
amplitude. The reason for this being that necessary condition for direct $CP$
-violation is that decay amplitude should be complex as discussed in section
1. But this is not sufficient because in the limit of no final state
interactions, the direct $CP$-violation in $B\rightarrow f$, $\bar{B}%
\rightarrow \bar{f}$ decay vanishes. To illustrate this point, we discuss
the decays $\bar{B}^{0}\rightarrow \pi ^{+}\pi ^{-}$. The main contribution
to this decay is from tree graph (see Fig. 4a); But this decay can also
proceed via the penguin diagram (see Fig. 4b).

The contribution of penguin diagram can be written as
\begin{equation}
P=V_{ub}V_{ud}^{\ast }f\left( u\right) +V_{cb}V_{cd}^{\ast }f\left( c\right)
+V_{tb}V_{td}^{\ast }f\left( t\right)  \label{4.1}
\end{equation}%
where $f\left( u\right) $, $f\left( c\right) $ and $f\left( t\right) $
denote the contributions of $u$, $c$ and $t$ quarks in the loop. Now using
the unitarity equation (\ref{07}), we can rewrite Eq. (\ref{4.1}) as,
\begin{eqnarray}
P_{c} &=&V_{ub}V_{ud}^{\ast }\left( f\left( u\right) -f\left( t\right)
\right) +V_{cb}V_{cd}^{\ast }\left( f\left( c\right) -f\left( t\right)
\right)  \notag \\
\text{or }P_{t} &=&V_{ub}V_{ud}^{\ast }\left( f\left( u\right) -f\left(
c\right) \right) +V_{tb}V_{td}^{\ast }\left( f\left( t\right) -f\left(
c\right) \right)  \notag \\
&&  \label{4.2}
\end{eqnarray}%
Due to loop integration $P$ is suppressed relative to $T$ but still its
contribution is not negligible. For the decay $\bar{B}^{0}\rightarrow f$ $%
(f=\pi ^{+}\pi ^{-})$\ the decay amplitude is given by
\begin{equation}
\bar{A}_{f}=A\left( \bar{B}^{0}\rightarrow \pi ^{+}\pi ^{-}\right)
=\left\vert T\right\vert e^{i\left( -\gamma +\delta _{T}\right) }+\left\vert
P\right\vert e^{i\left( \phi +\delta _{P}\right) }  \label{4.3}
\end{equation}%
where $\delta _{T}$ and $\delta _{P}$ are strong interaction phases which
have been taken out $\phi $ is the weak phase in Penguin graph. $CPT$
invariance gives,%
\begin{equation}
A_{f}\equiv {}A\left( B^{0}\rightarrow \pi ^{+}\pi ^{-}\right) =\left\vert
T\right\vert e^{-i\left( -\gamma -\delta _{T}\right) }+\left\vert
P\right\vert e^{-i\left( \phi -\delta _{P}\right) }.  \label{4.4}
\end{equation}%
Direct $CP$--violation asymmetry is given by,
\begin{eqnarray}
A_{CP} &=&\frac{-\Gamma \left( B^{0}\rightarrow \pi ^{+}\pi ^{-}\right)
+\Gamma \left( \bar{B}^{0}\rightarrow \pi ^{+}\pi ^{-}\right) }{\Gamma
\left( B^{0}\rightarrow \pi ^{+}\pi ^{-}\right) +\Gamma \left( \bar{B}%
^{0}\rightarrow \pi ^{+}\pi ^{-}\right) }  \notag \\
&=&-\frac{1-\left\vert \lambda \right\vert ^{2}}{1+\left\vert \lambda
\right\vert ^{2}}=-C_{\pi \pi }
\end{eqnarray}%
For the second choice,
\begin{equation}
\phi =\beta ,\ F_{\text{CKM}}=\frac{\left\vert V_{tb}\right\vert \left\vert
V_{td}\right\vert }{\left\vert V_{ub}\right\vert \left\vert
V_{ud}\right\vert }\approx \frac{\sqrt{\left( 1-\bar{\rho}\right) ^{2}+\bar{%
\eta}^{2}}}{\sqrt{\bar{\rho}^{2}+\bar{\eta ^{2}}}},\text{ }r=\frac{R_{t}}{%
R_{b}}\frac{\left\vert P_{t}\right\vert }{\left\vert T\right\vert }
\label{4.7a}
\end{equation}%
\begin{eqnarray}
A_{\pi ^{+}\pi ^{-}} &=&\left\vert T\right\vert e^{i\gamma }e^{i\delta
_{T}}[1-re^{i(\alpha +\delta )}]  \label{4.8} \\
\delta &=&\delta _{P}-\delta _{T}
\end{eqnarray}%
Hence we get, from Eqs (\ref{3.14}), (\ref{3.14b}), (\ref{93}) and (\ref{4.8}%
)%
\begin{eqnarray}
C_{\pi ^{+}\pi ^{-}} &=&-A_{CP}=\frac{2r\sin \delta \sin \alpha }{1-2r\cos
\delta \cos \alpha +r^{2}} \\
S_{\pi ^{+}\pi ^{-}} &=&\frac{\sin 2\alpha -2r\cos \delta \sin \alpha }{%
1-2r\cos \delta \cos \alpha +r^{2}}\approx \sin 2\alpha +2r\cos \delta \sin
\alpha \cos 2\alpha  \notag
\end{eqnarray}

\begin{figure}[tbp]
\begin{center}
\includegraphics[scale=0.4]{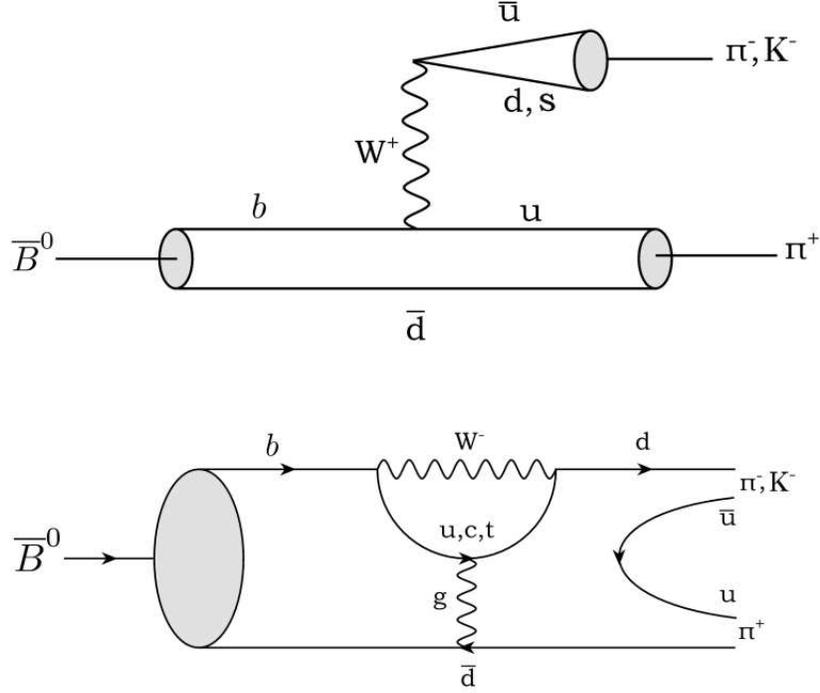}
\end{center}
\caption{Tree and Penguin Diagrams}
\end{figure}

From Eqs. (\ref{3.14b}), (\ref{4.7a}) and (\ref{4.8}), we can express $%
S_{\pi ^{+}\pi ^{-}}$\ in the form%
\begin{eqnarray}
S_{\pi ^{+}\pi ^{-}} &=&\sqrt{1-C_{\pi ^{+}\pi ^{-}}^{2}}\frac{1}{|\lambda |}%
\text{Im}[e^{-2i\beta }\lambda ]  \notag \\
&=&\sqrt{1-C_{\pi ^{+}\pi ^{-}}^{2}}\text{Im}[e^{-2i\beta }e^{-2i\gamma
}e^{i\delta _{+-}}]  \notag \\
&=&\sqrt{1-C_{\pi ^{+}\pi ^{-}}^{2}}\sin [2\alpha +\delta _{+-}]
\end{eqnarray}%
where we have put
\begin{subequations}
\begin{equation}
\lambda =\frac{\overline{A}_{f}}{A_{f}}=\frac{|\overline{A}_{f}|}{\left\vert
A_{f}\right\vert }e^{-2i\gamma }e^{i\delta _{+-}}
\end{equation}%
For $B^{0}\rightarrow \pi ^{0}\pi ^{0}$,$B^{+}\rightarrow \pi ^{+}\pi ^{0}$
the decay amplitudes are given by,
\end{subequations}
\begin{eqnarray}
A_{00} &=&A(B^{0}\rightarrow \pi ^{0}\pi ^{0})=\frac{1}{\sqrt{2}}\left\vert
T\right\vert e^{i\delta _{T}}e^{i\gamma }\left[ -r_{c}e^{i\delta
_{CT}}-re^{i(\alpha +\delta )}\right]  \notag \\
A_{+0} &=&A(B^{+}\rightarrow \pi ^{+}\pi ^{0})=\frac{1}{\sqrt{2}}\left\vert
T\right\vert e^{i\delta _{T}}e^{i\gamma }\left[ 1+r_{C}e^{i\delta _{CT}}%
\right]  \notag \\
r_{C} &=&\frac{\left\vert C\right\vert }{\left\vert T\right\vert },\qquad
\delta _{CT}=\delta _{C}-\delta _{T},
\end{eqnarray}%
Thus%
\begin{eqnarray}
A_{+0}^{CP} &=&0  \notag \\
C_{\pi ^{0}\pi ^{0}} &=&-A_{00}^{CP}=\frac{-2r/r_{C}\sin (\delta -\delta
_{CT})\sin \alpha }{1+r^{2}/r_{C}^{2}+2r/r_{C}\cos (\delta -\delta
_{CT})\cos \alpha }  \notag \\
&& \\
S_{\pi ^{0}\pi ^{0}} &=&\frac{\sin 2\alpha +2r/r_{C}\cos (\delta -\delta
_{CT})\sin \alpha }{1+r^{2}/r_{C}^{2}+2r/r_{C}\cos (\delta -\delta
_{CT})\cos \alpha }
\end{eqnarray}%
Experimental values for $CP$-asymmetries are%
\begin{equation*}
C_{\pi ^{+}\pi ^{-}}=0.38\pm 0.17,\text{ }S_{\pi ^{+}\pi ^{-}}=-0.61\pm 0.08
\end{equation*}%
For $B^{0}(\overline{B}^{0})\rightarrow K^{+}\pi ^{-}(K^{-}\pi ^{+})$ decay $%
CP\left\vert f\right\rangle =\left\vert \overline{f}\right\rangle \neq
\left\vert f\right\rangle $. For this case only direct CP-violation is
possible. It is easy to see that the decay amplitude can be expressed in the
following form%
\begin{eqnarray*}
A(\overline{B}^{0} &\rightarrow &K^{-}\pi ^{+})=T+P=|P|e^{i\delta
_{P}}[1+re^{i(-\gamma -\delta )}] \\
A(B^{0} &\rightarrow &K^{+}\pi ^{-})=|P|e^{i\delta _{P}}[1+re^{i(\gamma
-\delta )}]
\end{eqnarray*}%
where%
\begin{eqnarray*}
\delta &=&\delta _{P}-\delta _{T} \\
r &=&\frac{|V_{ub}||V_{us}|}{|V_{cb}||V_{cs}|}\frac{|T|}{|P|}
\end{eqnarray*}%
Hence%
\begin{equation*}
A_{CP}(K^{+}\pi ^{-})=\frac{-2r\sin \gamma \sin \delta }{1+2r\cos \gamma
\cos \delta +r^{2}}=-0.089\pm 0.013\text{ (Expt value)}
\end{equation*}

Finally it is convenient to write, from Eqs.(\ref{I}) and (\ref{II}), the
decay rates in the following form,
\begin{eqnarray}
&&\left[ \Gamma _{f}(t)-\bar{\Gamma}_{\bar{f}}(t)\right] +\left[ \Gamma _{%
\bar{f}}-\bar{\Gamma}_{f}(t)\right]  \notag \\
&=&e^{-\Gamma t}\left\{ \cos \Delta mt\left[ \left( \left\vert
A_{f}\right\vert ^{2}-\left\vert \bar{A}_{\bar{f}}\right\vert ^{2}\right)
+\left( \left\vert A_{\bar{f}}\right\vert ^{2}-\left\vert \bar{A}%
_{f}\right\vert ^{2}\right) \right] \right.  \notag \\
&&\left. +2\sin \Delta mt\left[ \text{Im}\left( e^{2i\phi _{M}}A_{f}^{\ast }%
\bar{A}_{f}\right) +\text{Im}\left( e^{2i\phi _{M}}A_{\bar{f}}^{\ast }\bar{A}%
_{\bar{f}}\right) \right] \right\}  \notag \\
&&  \label{e1} \\
&&\left[ \Gamma _{f}(t)+\bar{\Gamma}_{\bar{f}}(t)\right] -\left[ \Gamma _{%
\bar{f}}(t)+\bar{\Gamma}_{f}(t)\right]  \notag \\
&=&e^{-\Gamma t}\left\{ \cos \Delta mt\left[ \left( \left\vert
A_{f}\right\vert ^{2}+\left\vert \bar{A}_{\bar{f}}\right\vert ^{2}\right)
-\left( \left\vert A_{\bar{f}}\right\vert ^{2}+\left\vert \bar{A}%
_{f}\right\vert ^{2}\right) \right] \right.  \notag \\
&&\left. +2\sin \Delta mt\left[ \text{Im}\left( e^{2i\phi _{M}}A_{f}^{\ast }%
\bar{A}_{f}\right) -\text{Im}\left( e^{2i\phi _{M}}A_{\bar{f}}^{\ast }\bar{A}%
_{\bar{f}}\right) \right] \right\}  \notag \\
&&  \label{e2}
\end{eqnarray}%
We end this section with the following remarks. The $CP$ asymmetries in the
hadronic decays of $B$, $B_{s}$\ and $K$\ mesons involve strong final state
phases. The strong interactions effects at the quark level are taken care of
by perturbative QCD in terms of Wilson coefficients. The CKM matrix which
connects the weak eigenstates with mass eigenstates is another aspect of
strong interactions at quark level. In the case of semi leptonic decays, the
long distance strong interaction effects manifest themselves in the form
factors of final states after hadronization. Likewise the strong interaction
final state phases are long distance effects. These phase shifts essentially
arise in terms of S-matrix which changes an 'in' state into an 'out' state
viz.
\begin{equation}
|f\rangle _{in}=S|f\rangle _{out}=e^{2i\delta _{f}}|f\rangle _{out}
\label{4.66}
\end{equation}%
In fact, the $CPT$ invariance of weak interaction Lagrangian gives for the
weak decay $B(\bar{B})\rightarrow f(\bar{f})$
\begin{equation}
\bar{A}_{\bar{f}}\equiv _{out}\langle \bar{f}|\mathcal{L}_{w}|\bar{B}\rangle
=\eta _{f}e^{2i\delta _{f}}A_{f}{\ast }  \label{4.67}
\end{equation}%
It is difficult to reliably estimate the final state strong phase shifts. It
involves the hadronic dynamics. However, using isospin, C-invariance of
S-matrix and unitarity of S-matrix, we can relate these phases. In this
regard, the decays $B^{0}\rightarrow f,\bar{f}$ described by two independent
single amplitudes $A_{f}$ and $A_{\bar{f}}^{\prime }$ discussed in section
6.2 and the decays described by the weak amplitudes $A_{f}\neq A_{\bar{f}}$,
described in section 6.3 are of interest

The $C$ invariance of S-matrix viz. $S_{\bar{f}}=S_{f}$ would imply
\begin{equation*}
\delta _{f}=\delta _{\bar{f}}^{\prime },\qquad \delta _{1f}=\delta _{1\bar{f}%
},\qquad \delta _{2f}=\delta _{2\bar{f}}
\end{equation*}%
In the above decays, b is converted into $b\rightarrow c(u)+\bar{u}+d$. In
particular, for the tree graph, the configuration is such that $\bar{u}$ and
d essentially go together into color singlet states while the third quark
c(u) recoiling; there is a significant probability that system will
hadronize as a two body final state. Thus at least for the tree amplitude $%
\delta _{f}^{T}=\delta _{\bar{f}}^{T}\approx 0$.

\subsection{Case II}

In this section we first consider the case in which single weak amplitudes $%
A_{f}$ and $A_{\bar{f}}^{^{\prime }}$ with different weak phases describe
the decays:
\begin{eqnarray}
A_{f} &=&\langle f\left\vert \mathcal{L_{W}}\right\vert B^{0}\rangle
=e^{i\phi }F_{f}  \notag \\
A_{\bar{f}} &\equiv &A_{\bar{f}}^{^{\prime }}=\langle \bar{f}\left\vert
\mathcal{L}^{\prime }\mathcal{_{W}}\right\vert B^{0}\rangle =e^{i\phi
^{^{\prime }}}F_{\bar{f}}^{^{\prime }}  \label{e3a}
\end{eqnarray}%
$CPT$ gives,
\begin{eqnarray}
\bar{A}_{\bar{f}} &=&\langle \bar{f}\left\vert \mathcal{L_{W}}\right\vert
\bar{B}^{0}\rangle =e^{2i\delta _{f}}A_{f}^{\ast }  \notag \\
\bar{A}_{f} &\equiv &\bar{A}_{f}^{^{\prime }}=\langle f\left\vert \mathcal{L}%
^{\prime }\mathcal{_{W}}\right\vert \bar{B}^{0}\rangle =e^{2i\delta _{\bar{f}%
}^{^{\prime }}}A_{\bar{f}}^{\ast ^{\prime }}  \label{e4}
\end{eqnarray}%
For these decays, only mixing induced $CP$-asymmetries are possible. Note $%
\delta _{f}$ and $\delta _{\bar{f}}^{\prime }$ are strong phases; $\phi $
and $\phi ^{\prime }$ are weak phases. The states $|f>$ and $|\overline{f}>$
are C-conjugate of each other such as states $D^{(\ast )-}\pi ^{+}(D^{(\ast
)+}\pi ^{-}),$ $D_{s}^{(\ast )-}K^{+}(D_{s}^{(\ast )+}K^{-}),$ $D^{-}\rho
^{+}(D^{+}\rho ^{-})$

For this case
\begin{eqnarray}
\mathcal{A}\left( t\right) &=&\frac{2\bigl|F_{f}\bigr|\bigl|F_{\bar{f}%
}^{\prime }\bigr|}{\bigl|F_{f}\bigr|^{2}+\bigl|F_{\bar{f}}\bigr|^{2}}\sin
\Delta mt\sin \bigl(2\phi _{M}-\phi -\phi \bigr)\cos \bigl(\delta
_{f}-\delta _{\bar{f}}^{\prime }\bigr)  \notag \\
&&  \label{e5} \\
\mathcal{F}\left( t\right) &=&\frac{\bigl|F_{f}\bigr|^{2}-\bigl|F_{\bar{f}%
}^{\prime }\bigr|^{2}}{\bigl|F_{f}\bigr|^{2}+\bigl|F_{\bar{f}}^{\prime }%
\bigr|^{2}}\cos \Delta mt  \notag \\
&&-\frac{2\bigl|F_{f}\bigr|\bigl|F_{\bar{f}}^{\prime }\bigr|}{\bigl|F_{f}%
\bigr|^{2}+\bigl|F_{\bar{f}}^{\prime }\bigr|^{2}}\sin \Delta mt\cos \left(
2\phi _{M}-\phi -\phi \right) \sin \bigl(\delta _{f}-\delta _{\bar{f}%
}^{\prime }\bigr)  \notag \\
&&  \label{e7}
\end{eqnarray}%
We now apply the above formula to $B\rightarrow \pi D$ and $B_{s}\rightarrow
KD_{s}$ decays. For these decays,
\begin{equation*}
\phi =0,\qquad \phi ^{\prime }=\gamma
\end{equation*}%
\begin{equation*}
\phi _{M}=%
\begin{cases}
-\beta , & \text{for $B^{0}$} \\
-\beta _{s}, & \text{for $B_{s}^{0}$}%
\end{cases}%
\end{equation*}%
Thus%
\begin{eqnarray*}
A_{f} &=&\langle D^{-}\pi ^{+}\left\vert \mathcal{L_{W}}\right\vert
B^{0}\rangle =F_{f} \\
A_{\bar{f}}^{\prime } &=&\langle D^{+}\pi ^{-}\left\vert \mathcal{L_{W}}%
^{\prime }\right\vert B^{0}\rangle =e^{i\gamma }F_{\bar{f}}^{^{\prime }} \\
A_{f_{s}} &=&\langle K^{+}D_{s}^{-}\left\vert \mathcal{L_{W}}\right\vert
B_{s}^{0}\rangle =F_{f_{s}} \\
A_{\bar{f}_{s}}^{\prime } &=&\langle K^{-}D_{s}^{+}\left\vert \mathcal{L_{W}}%
^{\prime }\right\vert B_{s}^{0}\rangle =e^{i\gamma }F_{\bar{f}%
_{s}}^{^{\prime }}
\end{eqnarray*}%
Note that the effective Lagrangians for decays $(q=d,s)$ are given by,
\begin{subequations}
\begin{eqnarray}
&&\mathcal{L_{W}}=V_{cb}V_{uq}^{\ast }\left[ \bar{q}\gamma ^{\mu }\left(
1-\gamma _{5}\right) u\right] \left[ \bar{c}\gamma _{\mu }\left( 1-\gamma
_{5}\right) b\right]  \label{e8a} \\
&&\mathcal{L_{W}}^{\prime }=V_{ub}V_{cq}^{\ast }\left[ \bar{q}\gamma ^{\mu
}\left( 1-\gamma _{5}\right) c\right] \left[ \bar{u}\gamma _{\mu }\left(
1-\gamma _{5}\right) b\right]  \label{e8b}
\end{eqnarray}%
respectively. In the Wolfenstein parametrization of $CKM$ matrix,
\end{subequations}
\begin{equation}
\frac{\left\vert V_{ub}\right\vert \left\vert V_{cq}\right\vert }{\left\vert
V_{cb}\right\vert \left\vert V_{uq}\right\vert }=\left( \lambda
^{2},1\right) R_{b},\qquad q=d,s  \label{e9}
\end{equation}%
Define,
\begin{equation*}
r=\lambda ^{2}R_{b}\frac{\bigl|F_{\bar{f}}^{\prime }\bigr|}{\bigl|F_{f}\bigr|%
},r_{s}=R_{b}\frac{\bigl|F_{\bar{f}_{s}}^{\prime }\bigr|}{\bigl|F_{f_{s}}%
\bigr|}
\end{equation*}%
Thus, we get from Eqs. $\eqref{e5}$ and $\eqref{e7}$ for $B^{0}$ decays,
(replacing $\frac{\bigl|F_{\bar{f}}^{\prime }\bigr|}{\bigl|F_{f}\bigr|}$ by $%
r$),%
\begin{eqnarray}
\mathcal{A}\left( t\right) &=&-\frac{2r}{1+r^{2}}\sin \Delta m_{B}t\sin
\left( 2\beta +\gamma \right) \cos \left( \delta _{f}-\delta _{\bar{f}%
}^{\prime }\right) \label{4.35ab}\\
\mathcal{F}\left( t\right) &=&\frac{1-r^{2}}{1+r^{2}}\cos \Delta m_{B}t-%
\frac{2r}{1+r^{2}}\sin \Delta m_{B}t\cos \left( 2\beta +\gamma \right) \sin
\left( \delta _{f}-\delta _{\bar{f}}^{\prime }\right)  \notag \\
&&  \label{4.35b}
\end{eqnarray}%
For the decays,
\begin{eqnarray*}
\bar{B}_{s}^{0}\left( B_{s}^{0}\right) &\rightarrow &K^{-}D_{s}^{+}\left(
K^{+}D_{s}^{-}\right) \\
\bar{B}_{s}^{0}\left( B_{s}^{0}\right) &\rightarrow &K^{+}D_{s}^{-}\left(
K^{-}D_{s}^{+}\right)
\end{eqnarray*}%
we get,
\begin{eqnarray}
\mathcal{A}_{s}\left( t\right) &=&-\frac{2r_{s}}{1+r_{s}^{2}}\sin (\Delta
m_{B_{s}}t)\sin \left( 2\beta _{s}+\gamma \right) \cos \left( \delta
_{f_{s}}-\delta _{\bar{f}_{s}}^{\prime }\right) \\
\mathcal{F}_{s}(t) &=&\frac{1-r_{s}^{2}}{1+r_{s}^{2}}\cos \Delta m_{B_{s}}t-%
\frac{2r_{s}}{1+r_{s}^{2}}\sin \Delta m_{B_{s}}t\cos \left( 2\beta
_{s}+\gamma \right) \sin \left( \delta _{f_{s}}-\delta _{\bar{f}%
_{s}}^{\prime }\right)  \notag \\
&&  \label{4.35}
\end{eqnarray}%
We note that for time integrated $CP$-asymmetry,
\begin{eqnarray}
\mathcal{A}_{s} &=&-\frac{2r_{s}}{1+r_{s}^{2}}\frac{\Delta m_{B_{s}}/\Gamma
_{s}}{1+\left( \Delta m_{B_{s}}/\Gamma _{s}\right) ^{2}}\sin \left( 2\beta
_{s}+\gamma \right) \cos (\delta _{f_{s}}-\delta _{\bar{f}_{s}}^{\prime })
\notag \\
&&  \label{4.35a}
\end{eqnarray}%
The $CP$--asymmetry $\mathcal{A}_{s}\left( t\right) $ or $\mathcal{A}_{s}$
involves two experimentally unknown parameters $\sin \left( 2\beta
_{s}+\gamma \right) $ and $\Delta m_{B_{s}}$. Both these parameters are of
importance in order to test the unitarity of $CKM$ matrix viz whether $CKM$
matrix is a sole source of $CP$--violation in the processes in which $CP$%
--violation has been observed.

From Eqs.$(\ref{4.35ab})$ and $(\ref{4.35b})$, we note that $CP$-asymetries:
\begin{eqnarray}
-\frac{S_{+}+S_{-}}{2} &=&\frac{2r}{1+r^{2}}\sin (2\beta +\gamma )\cos
(\delta _{f}-\delta _{\overline{f}}^{\prime }) \\
-\frac{S_{+}-S_{-}}{2} &=&\frac{2r}{1+r^{2}}\cos (2\beta +\gamma )\sin
(\delta _{f}-\delta _{\overline{f}}^{\prime })
\end{eqnarray}%
involve the weak phase $2\beta +\gamma $ and strong phase $\delta
_{f}-\delta _{\overline{f}}^{\prime }$. For $B_{s}^{0},$ replace $%
r\rightarrow r_{s}$, $\delta _{f}\rightarrow \delta _{f_{s}}$, $\delta _{%
\overline{f}}^{\prime }=\delta _{\overline{f}_{s}}^{\prime }$ and $\beta $
by $\beta _{s}$. In the standared model $\beta _{s}=0$. The decays $%
B\rightarrow \pi D,\pi D^{\ast },\rho D$\ and $B_{s}\rightarrow KD_{s},\pi
D_{s}^{\ast },K^{\ast }D_{s}$ are described by tree amplitudes (see Fig 5).
For tree graphs, we assume factorization, factorization implies $\delta
_{f}=\delta _{\overline{f}}^{\prime }=0$ hence we have%
\begin{eqnarray}
-\frac{S_{+}+S_{-}}{2} &=&\frac{2r}{1+r^{2}}\sin (2\beta +\gamma ) \\
-\frac{S_{+}-S_{-}}{2} &=&0  \label{c0}
\end{eqnarray}%
The experimental values for these asymmetries are
\begin{eqnarray}
\frac{S_{+}+S_{-}}{2} &=&\left[
\begin{array}{c}
-0.037\pm 0.012\text{ \ \ }B^{0}\rightarrow D^{\ast -}\pi ^{+} \\
-0.046\pm 0.023\text{ \ \ }B^{0}\rightarrow D^{-}\pi ^{+}%
\end{array}\right.
\\
\frac{S_{+}-S_{-}}{2} &=&\left[
\begin{array}{c}
-0.006\pm 0.016\text{ \ \ }B^{0}\rightarrow D^{\ast -}\pi ^{+} \\
-0.022\pm 0.021\text{ \ \ }B^{0}\rightarrow D^{-}\pi ^{+}%
\end{array}\right.
\end{eqnarray}%
Equation (\ref{c0}) is consistent with experimental values. Factorization
gives for the decay $\bar{B}^{0}\rightarrow D^{(\ast )+}\pi ^{-}$:
\begin{align}
|\bar{F}_{\bar{f}}|=|\bar{T}_{\bar{f}}|& =G[f_{\pi
}(m_{B}^{2}-m_{D}^{2})f_{0}^{B-D}(m_{\pi }^{2}),2f_{\pi }m_{B}|\vec{p}%
|A_{0}^{B-D^{\ast }}(m_{\pi }^{2}),  \label{c1} \\
|\bar{F}_{f}^{^{\prime }}|=|\bar{T}_{f}^{^{\prime }}|& =G^{^{\prime
}}[f_{D}(m_{B}^{2}-m_{\pi }^{2})f_{0}^{B-\pi }(m_{D}^{2}),2f_{D^{\ast
}}m_{B}|\vec{p}|f_{+}^{B-\pi }(m_{D^{\ast }}^{2}),  \label{c2} \\
G& =\frac{G_{F}}{\sqrt{2}}|V_{ud}||V_{cb}|a_{1},\quad G^{^{\prime }}=\frac{%
G_{F}}{\sqrt{2}}|V_{cd}||V_{ub}|  \label{c3}
\end{align}%
From the experimental branching ratios, the form factors $A_{0}^{B-D}(m_{\pi
}^{2})$ and $f_{0}^{B-D}(m_{\pi }^{2})$ can be obtained. In terms of the
form factors in HQET,%
\begin{eqnarray*}
f_{0}(t) &=&\frac{\sqrt{m_{B}m_{D}}}{m_{B}+m_{D}}(1+\omega )h_{0}(\omega ) \\
A_{0}(t) &=&\frac{m_{B}+m_{D^{\ast }}}{2\sqrt{m_{B}m_{D^{\ast }}}}%
h_{A_{0}}(\omega ) \\
A(t) &=&\frac{\sqrt{m_{B}m_{D^{\ast }}}}{m_{B}+m_{D^{\ast }}}(1+\omega
)h_{A_{1}}(\omega ) \\
t &=&m_{B}^{2}+m_{D^{\ast }}^{2}-2m_{B}m_{D^{\ast }}
\end{eqnarray*}

\begin{figure}[tbp]
\begin{center}
\includegraphics[scale=0.5]{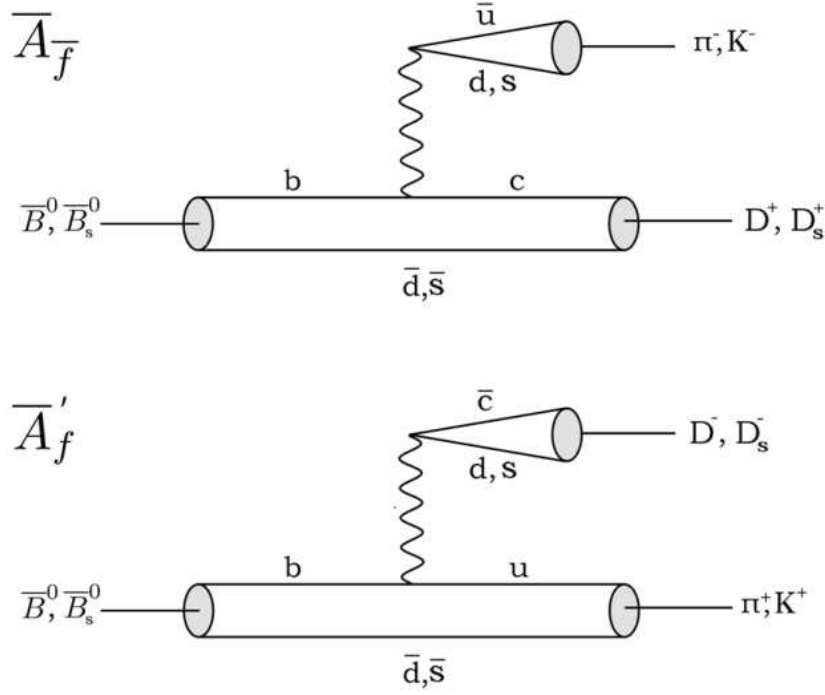}
\end{center}
\caption{a) Tree diagrams for $\overline{B}^{0}(\overline{B}%
_{s}^{0})\rightarrow D^{+}\protect\pi ^{-}(D_{s}^{+}K^{-})$ b) Tree diagrams
for $\overline{B}^{0}(\overline{B}_{s}^{0})\rightarrow D^{-}\protect\pi %
^{+}(D_{s}^{-}K^{+})$}
\end{figure}

we get%
\begin{equation}
h_{0}(\omega _{\max }^{{}})=0.51\pm 0.03,\text{ \ \ \ \ \ }h_{A_{0}}(\omega
_{\max }^{\ast })=0.54\pm 0.03
\end{equation}%
to be compared with the value%
\begin{equation}
|h_{A_{1}}(\omega _{max}^{\ast })|=0.52\pm 0.03  \label{c10}
\end{equation}%
obtained from the analysis of semi-leptonic decay $\overline{B}%
^{0}\rightarrow D^{(\ast )+}l^{-}\overline{\nu }_{l}$. The agreement between
the wo values is remarkable. Hence the factorization assumption for $%
B^{0}\rightarrow \pi D^{(\ast )}$ decays is experimentally on solid footing
and is in agreement with HQET.

From Eqs. $\eqref{c1}$ and $\eqref{c2}$, we obtain
\begin{align}
r& =\lambda ^{2}R_{b}\frac{|\bar{T}_{f}^{^{\prime }}|}{|\bar{T}_{\bar{f}}|}
\notag \\
& =\lambda ^{2}R_{b}\left[ \frac{f_{D}(m_{B}^{2}-m_{\pi }^{2})f_{0}^{B-\pi
}(m_{D}^{2})}{f_{\pi }(m_{B}^{2}-m_{D}^{2})f_{0}^{B-D}(m_{\pi }^{2})},\quad
\frac{f_{D^{\ast }}f_{+}^{B-\pi }(m_{D^{\ast }}^{2})}{f_{\pi
}A_{0}^{B-D^{\ast }}(m_{\pi }^{2})}\right]  \label{c11}
\end{align}%
Using the value of $r$ obtained from $\eqref{c11}$ one gets
\begin{equation}
-\left( \frac{S_{+}+S_{-}}{2}\right) _{D^{\ast }\pi }=2(0.017\pm 0.003)\sin
(2\beta +\gamma )  \label{c15}
\end{equation}%
Using the experimental value of the $CP$ asymmetry for $B^{0}\rightarrow
D^{\ast }\pi $ decay which has the least error, one gets the following
bounds
\begin{align}
\sin (2\beta +\gamma )& >0.69  \label{c16} \\
44^{\circ }& \leq (2\beta +\gamma )\leq 90^{\circ } \\
\text{or}\quad 90^{\circ }& \leq (2\beta +\gamma )\leq 136^{\circ }
\end{align}%
Selecting the second solution, and using $2\beta \approx 43^{\circ }$, we
get
\begin{equation}
\gamma =(70\pm 23)^{\circ }  \label{c19}
\end{equation}%
To end this section, we discuss the decays $\bar{B}_{s}^{0}\rightarrow
D_{s}^{+}K^{-},D_{s}^{\ast +}K^{-}$ are for which no experimental data are
available. Howevere using factorization and SU(3) one gets the following
branching ratios
\begin{equation}
\frac{\Gamma (\bar{B_{s}}^{0}\rightarrow D_{s}^{(\ast )+}K^{-})}{\Gamma _{%
\bar{B}_{s}^{0}}}=(1.94\pm 0.07)\times 10^{-4}[(1.96\pm 0.07)\times 10^{-4}]
\label{c34}
\end{equation}%
and
\begin{equation}
-\left( \frac{S_{+}+S_{-}}{2}\right) _{D_{s}^{\ast }K}=(0.41\pm 0.08)\sin
(2\beta _{s}+\gamma )
\end{equation}%
In the standard model with three generations of quarks $\beta _{s}=0.$ This
asymmetry is of spacial interest to test the physics beyond standard model.
The experimental results for these\ decays will be relevant not only for the
standard model but also for physics beyond standard model.

\subsection{Case III: $A_{f}\neq A_{\bar{f}}$}

\begin{eqnarray*}
A_{f} &=&\langle f\left\vert \mathcal{L_{W}}\right\vert B^{0}\rangle =\left[
e^{i\phi _{1}}F_{1f}+e^{i\phi _{2}}F_{2f}\right] \\
A_{\bar{f}} &=&\langle \bar{f}\left\vert \mathcal{L_{W}}\right\vert
B^{0}\rangle =\left[ e^{i\phi _{1}}F_{1\bar{f}}+e^{i\phi _{2}}F_{2\bar{f}}%
\right]
\end{eqnarray*}%
Examples:
\begin{equation*}
B^{0}\rightarrow \rho ^{-}\pi ^{+}(f):\text{ }A_{f}\qquad B^{0}\rightarrow
\rho ^{+}\pi ^{-}(\bar{f}):A_{\bar{f}}
\end{equation*}%
\begin{equation*}
B_{s}^{0}\rightarrow K^{\ast -}K^{+}\qquad B_{s}^{0}\rightarrow K^{\ast
+}K^{-}
\end{equation*}%
$CPT$ gives,
\begin{equation*}
\bar{A}_{\bar{f},f}=\sum_{i}[e^{-i\phi _{i}}e^{2i\delta _{f,\bar{f}%
}^{i}}F_{if\bar{f}}^{\ast }]
\end{equation*}%
For these decays, subtracting and adding Eqs. $(\ref{e1})$ and $(\ref{e2})$,
we get,%
\begin{align}
\frac{\Gamma _{\bar{f}}(t)-\bar{\Gamma}_{\bar{f}}(t)}{\Gamma _{\bar{f}}(t)+%
\bar{\Gamma}_{\bar{f}}(t)}=& C_{\bar{f}}\cos \Delta mt+S_{\bar{f}}\sin
\Delta mt  \label{ccc1} \\
=& (C+\Delta C)\cos \Delta mt+(S+\Delta S)\sin \Delta mt  \notag
\end{align}%
\begin{align}
\frac{\Gamma _{f}(t)-\bar{\Gamma}_{f}(t)}{\Gamma _{f}(t)+\bar{\Gamma}_{f}(t)}%
=& C_{f}\cos \Delta mt+S_{f}\sin \Delta mt  \notag \\
=& (C-\Delta C)\cos \Delta mt+(S-\Delta S)\sin \Delta mt  \label{ccc2}
\end{align}%
For these decays, the decay amplitudes can be written in terms of tree
amplitude $e^{i\phi _{T}}T_{f}$ and the penguin amplitude $e^{i\phi
_{P}}P_{f}$. We confine to decays:%
\begin{equation}
B^{0}\rightarrow \rho ^{-}\pi ^{+}:A_{f};\qquad B^{0}\rightarrow \rho
^{+}\pi ^{-}:A_{\bar{f}};\quad \phi _{T}=\gamma ,\phi _{P}=-\beta
\label{ccc14}
\end{equation}%
Hence for $B^{0}\rightarrow \rho ^{-}\pi ^{+},\overline{B}^{0}\rightarrow
\rho ^{+}\pi ^{-},B^{0}\rightarrow \rho ^{+}\pi ^{-},\overline{B}%
^{0}\rightarrow \rho ^{-}\pi ^{+}$, we have
\begin{align}
A_{f}& =\bigl|T_{f}\bigr|e^{-i\gamma }e^{i\delta
_{f}^{T}}[1-r_{f}e^{i(\alpha +\delta _{f})}]  \notag \\
A_{\bar{f}}& =\bigl|T_{\bar{f}}\bigr|e^{-i\gamma }e^{i\delta _{\bar{f}%
}^{T}}[1-r_{\bar{f}}e^{i(\alpha +\delta _{\bar{f}})}]  \label{ccc16} \\
\text{where}\qquad r_{f,\bar{f}}& =\frac{|V_{tb}||V_{td}|}{|V_{ub}||V_{ud}|}%
\frac{\bigl|P_{f,\bar{f}}\bigr|}{\bigl|T_{f,\bar{f}}\bigr|}=\frac{R_{t}}{%
R_{b}}\frac{\bigl|P_{f,\bar{f}}\bigr|}{\bigl|T_{f,\bar{f}}\bigr|}
\label{ccc17}
\end{align}%
In order to take into account the contributions of tree and penguin diagram,
we introduce the angles $\alpha _{eff}^{f,\bar{f}}$ , defined as follows
\begin{align}
e^{i\beta }A_{f,\bar{f}}& =|A_{f,\bar{f}}|e^{-i\alpha _{eff}^{f,\bar{f}}}
\notag \\
e^{-i\beta }\bar{A}_{\bar{f},f}& =|\bar{A}_{\bar{f},f}|e^{i\alpha _{eff}^{f,%
\bar{f}}}  \label{ccc34}
\end{align}%
With this definition, we separate out tree and penguin contributions:
\begin{eqnarray}
e^{i\beta }A_{f,\bar{f}}-e^{-i\beta }\bar{A}_{\bar{f},f}& =&|A_{f,\bar{f}%
}|e^{-i\alpha ^{f,\bar{f}}}-|\bar{A}_{\bar{f},f}|e^{i\alpha ^{f,\bar{f}}}
\notag \\
& =&2iT_{f,\bar{f}}\sin \alpha  \label{ccc35} \\
e^{i(\alpha +\beta )}A_{f,\bar{f}}-e^{-i(\alpha +\beta )}\bar{A}_{\bar{f}%
,f}& =&|A_{f,\bar{f}}|e^{-i(\alpha _{eff}^{f,\bar{f}}-\alpha )}-|A_{\bar{f}%
,f}|e^{i(\alpha _{eff}^{f,\bar{f}}-\alpha )} \nonumber \\
& =&(2iT_{f,\bar{f}}\sin \alpha )r_{f,\bar{f}}e^{i\delta _{f,\bar{f}}}  \nonumber
\\
& =&2iP_{f,\bar{f}}\sin \alpha  \label{ccc36}
\end{eqnarray}%
From Eqs. $\eqref{ccc35}$ and $\eqref{ccc36}$, we get%
\begin{equation}
R_{f,\bar{f}}\left[ 1-\sqrt{1-A_{CP}^{f,\bar{f}2}}\cos (2\alpha _{eff}^{f,%
\bar{f}})\right] =2\left\vert T_{f,\bar{f}}\right\vert ^{2}\sin ^{2}\alpha
\end{equation}%
\begin{eqnarray}
r_{f,\bar{f}}^{2}& =&\frac{1-\sqrt{1-A_{CP}^{f,\bar{f}2}}\cos (2\alpha
_{eff}^{f,\bar{f}}-2\alpha )}{1-\sqrt{1-A_{CP}^{f,\bar{f}2}}\cos 2\alpha
_{eff}^{f,\bar{f}}}  \label{ccc39} \\
r_{f,\bar{f}}\cos \delta _{f,\bar{f}}& =&\frac{\cos \alpha -\sqrt{1-A_{CP}^{f,
\bar{f}2}}\cos (2\alpha _{eff}^{f,\bar{f}}-\alpha )}{1-\sqrt{1-A_{CP}^{f,
\bar{f}2}}\cos 2\alpha _{eff}^{f,\bar{f}}}  \label{ccc40} \\
r_{f,\bar{f}}\sin \delta _{f,\bar{f}}& =&\frac{-A_{CP}^{f,\bar{f}}\sin \alpha
}{1-\sqrt{1-A_{CP}^{f,\bar{f}2}}\cos 2\alpha _{eff}^{f,\bar{f}}}
\label{ccc41} \\
S_{f,\bar{f}}& =&\sqrt{1-C_{f,\bar{f}}^{2}}\sin (2\alpha _{eff}^{f,\bar{f}
}\mp \delta )
\end{eqnarray}%
where the phase $\delta $\ is defined as
\begin{equation}
\bar{A}_{\bar{f}}=\frac{|\bar{A}_{\bar{f}|}}{|\bar{A}_{f}|}\bar{A}%
_{f}e^{i\delta }  \label{cccc9}
\end{equation}%
Thus one sees it is convenient to analyse these decays in terms of $\alpha
_{eff}^{f,\bar{f}}$. From Eq. $\eqref{ccc35}$, we get
\begin{align}
\sin 2\delta _{f,\bar{f}}^{T}& =-A_{CP}^{f,\bar{f}}\frac{\sin 2\alpha
_{eff}^{f,\bar{f}}}{1-\sqrt{1-A_{CP}^{f,\bar{f}2}}\cos 2\alpha _{eff}^{f,%
\bar{f}}}  \label{ccc38a} \\
\cos 2\delta _{f,\bar{f}}^{T}& =\frac{\sqrt{1-A_{CP}^{f,\bar{f}2}}-\cos
2\alpha _{eff}^{f,\bar{f}}}{1-\sqrt{1-A_{CP}^{f,\bar{f}2}}\cos 2\alpha
_{eff}^{f,\bar{f}}}
\end{align}%
Now factorization implies
\begin{equation}
\delta _{f}^{T}=0=\delta _{\bar{f}}^{T}  \label{ccc42}
\end{equation}%
Thus in the limit $\delta _{f}^{T}\rightarrow 0$, we get from Eqs. (\ref%
{ccc38a})
\begin{equation}
\cos 2\alpha _{eff}^{f,\bar{f}}=-1,\qquad \alpha _{eff}^{f,\bar{f}%
}=90^{\circ }  \label{ccc43}
\end{equation}%
and from Eqs. (\ref{ccc40}), (\ref{ccc41}) and (\ref{ccc43})%
\begin{align}
r_{f,\bar{f}}\cos \delta _{f,\bar{f}}& =\cos \alpha  \label{ccc44} \\
r_{f,\bar{f}}\sin \delta _{f,\bar{f}}& =\frac{-A_{CP}^{f,\bar{f}}\sin \alpha
}{1+\sqrt{1-A_{CP}^{f,\bar{f}2}}}  \label{ccc45} \\
r_{f,\bar{f}}^{2}& =\frac{1+\sqrt{1-A_{CP}^{f,\bar{f}2}}\cos 2\alpha }{1+%
\sqrt{1-A_{CP}^{f,\bar{f}2}}}  \label{ccc48}
\end{align}%
Finally the $CP$ asymmetries in the limit $\delta _{f,\bar{f}%
}^{T}\rightarrow 0$
\begin{align}
S_{\bar{f}}=S+\Delta S& =-\sqrt{1-C_{\bar{f}}^{2}}\cos \delta \\
S_{f}=S-\Delta S& =\sqrt{1-C_{f}^{2}}\cos \delta
\end{align}%
For $B^{0}(\bar{B}^{0})\rightarrow \rho ^{-}\pi ^{+},\rho ^{+}\pi ^{-}(\rho
^{+}\pi ^{-},\rho ^{-},\pi ^{+})$ decays the experimental results are
\begin{align}
\Gamma & =R_{f}+R_{\bar{f}}=(22.8\pm 2.5)\times 10^{-6}  \label{ccc19} \\
A_{CP}^{f}& =-0.16\pm 0.23,\quad A_{CP}^{\bar{f}}=0.08\pm 0.12  \label{ccc20}
\\
C& =0.01\pm 0.14,\quad \Delta C=0.37\pm 0.08  \label{ccc21} \\
S& =0.01\pm 0.09,\quad \Delta S=-0.05\pm 0.10  \label{ccc22}
\end{align}%
With above values, it is hard to draw any reliable conclusion.

\section{$CP$-Violation in Hadronic Weak Decays of Baryons}

So far we have discussed the $CP$ violation in $K^{0}-\overline{K}%
^{0},B_{q}^{0}-\overline{B}_{q}^{0}$\ systems. There is a need to study $CP$
violation outside these systems. The hadronic weak decays of baryons and
antibaryons provide another framework to study $CP$ violation.

The hadronic weak decays%
\begin{equation*}
N(p)\rightarrow N(p^{\prime })+\pi (q)
\end{equation*}%
is described by the amplitude
\begin{equation}
M_{f}=\overline{u}(\mathbf{p}^{\prime })[A-\gamma _{5}B]u(\mathbf{p})\sim
\chi ^{\dag }[a_{s}+a_{p}\mathbf{\sigma }.\mathbf{n}]\chi
\end{equation}%
(Note here we have designated a baryon by N, not to confuse with a B-meson
and $\pi $\ is any pseudoscalar meson).\newline
Under charge conjugation (\textit{C}):%
\begin{equation*}
u(\mathbf{p})\rightarrow C\overline{v}^{T}(\mathbf{p}),\ \ C=i\gamma
^{0}\gamma ^{2}
\end{equation*}%
Under space reflection (\textit{P}):%
\begin{equation*}
u^{(r)}(\mathbf{p})\rightarrow u^{(r)}(-\mathbf{p})=\gamma ^{0}u(\mathbf{p})
\end{equation*}%
Under time reversal (\textit{T}):%
\begin{equation*}
u^{(r)}(\mathbf{p})\rightarrow u^{\ast (-r)}(-\mathbf{p})=Bu^{(r)}(\mathbf{p}%
),\ \ B=\gamma ^{1}\gamma ^{3}
\end{equation*}%
Thus, under these transformations%
\begin{equation*}
M_{f}\overset{CP}{\rightarrow }-\overline{v}(\mathbf{p})[A+\gamma _{5}B]v(%
\mathbf{p}^{\prime })=\overline{M}\overline{_{f}}\sim \chi ^{\dag }\left(
-a_{s}+a_{p}\mathbf{\sigma }.\mathbf{n}\right)
\end{equation*}%
\begin{equation*}
M_{f}\overset{T}{\rightarrow }\overline{u}(\mathbf{p}^{\prime })[A^{\ast
}-\gamma _{5}B^{\ast }]u(\mathbf{p})
\end{equation*}%
\begin{equation}
M_{f}\overset{CPT}{\rightarrow }-\overline{v}(\mathbf{p})[A^{\ast }+\gamma
_{5}B^{\ast }]v(\mathbf{p}^{\prime })=\overline{M}\overline{_{f}}
\end{equation}%
When final state interactions are taken into account, the partial wave
amplitudes $a_{s}$\ and $a_{p}$\ acquire sttrong final state phases $%
e^{i\delta _{f}^{s}}$\ and $e^{i\delta _{f}^{p}}$\ respectively. Thus with
final state interactions
\begin{equation*}
\underset{out}{}\left\langle f\left\vert H\right\vert B\right\rangle \overset%
{CPT}{\rightarrow }\underset{in}{}\left\langle \overline{f}\left\vert
H\right\vert \overline{B}\right\rangle ^{\ast }=e^{2i\delta _{f}}\underset{%
out}{}\left\langle \overline{f}\left\vert H\right\vert \overline{B}%
\right\rangle ^{\ast }
\end{equation*}%
Hence under $CP$ and $CPT$%
\begin{subequations}
\begin{eqnarray}
M_{f}\overset{CP}{\rightarrow }\overline{M}\overline{_{f}} &=&\chi ^{\dag
}[-\left\vert a_{s}\right\vert e^{i\delta _{f}^{s}}+\left\vert
a_{p}\right\vert e^{i\delta _{f}^{p}}\mathbf{\sigma }.\mathbf{n}]\chi
\label{5.1} \\
&&M_{f}\overset{CPT}{\rightarrow }-\overline{v}(\mathbf{p}^{\prime
})[e^{2i\delta _{f}^{s}}A^{\ast }-\gamma _{5}e^{2i\delta _{f}^{p}}B^{\ast
}]v(\mathbf{p})  \notag \\
&=&\overline{M}\overline{_{f}}\sim \chi ^{\dag }[-e^{2i\delta
_{f}^{s}}a_{s}^{\ast }+e^{2i\delta _{f}^{p}}a_{p}^{\ast }\mathbf{\sigma }.%
\mathbf{n}]\chi  \label{5.2}
\end{eqnarray}%
Hence from Eq (\ref{5.1}), we conclude that $CP$ symmetry gives
\end{subequations}
\begin{equation}
\overline{\Gamma }=\Gamma ,\text{ \ \ \ \ }\overline{\alpha }=\alpha ,\text{
\ \ \ \ }\overline{\beta }=\beta  \label{5.3}
\end{equation}%
From Eqs (\ref{5.1}) and (\ref{5.2}), we note that both $CP$ and $CPT$
invariance give the same result given in Eq (\ref{5.3}), unless the S-wave
amplitude A and P-wave amplitude B have different weak phases. Hence to
leading order, $CP$-odd observables%
\begin{eqnarray*}
\delta \Gamma &=&\frac{\Gamma -\overline{\Gamma }}{\Gamma +\overline{\Gamma }%
} \\
\delta \alpha &=&\frac{\alpha +\overline{\alpha }}{\alpha -\overline{\alpha }%
} \\
\delta \beta &=&\frac{\beta +\overline{\beta }}{\beta -\overline{\beta }}
\end{eqnarray*}%
are non zero only if the above condition is satisfied.

The decays of $B(\bar{B})$ mesons to baryon-antibaryon pair $N_{1}$ $\bar{N}%
_{2}$ $(\bar{N}_{1}$ $N_{2})$ and subsequent decays of $N_{2},\bar{N}_{2}$
or $(N_{1},\bar{N}_{1})$ to a lighter hyperon (antihyperon) plus a meson
also provide a means to study $CP$-odd observables as for example in the
process,
\begin{equation*}
e^{-}e^{+}\rightarrow B,\bar{B}\rightarrow N_{1}\bar{N}_{2}\rightarrow N_{1}%
\bar{N}_{2}^{\prime }\bar{\pi},\qquad \bar{N}_{1}N_{2}\rightarrow \bar{N}%
_{1}N_{2}^{\prime }\pi
\end{equation*}%
The decay $B\rightarrow N_{1}\bar{N}_{2}(f)$ is described by the matrix
element,
\begin{equation}
M_{f}=F_{q}e^{+i\phi }\left[ \bar{u}(\mathbf{p}_{1})(A_{f}+\gamma
_{5}B_{f})v(\mathbf{p}_{2})\right]  \label{q1}
\end{equation}%
where as $B\rightarrow \overline{N}_{1}N_{2}(\overline{f})$ is described by
the matrix elements
\begin{equation}
\overset{}{M^{\prime }}_{f}=\overset{}{F^{\prime }}_{q}e^{+i\phi ^{^{\prime
}}}\left[ \bar{u}(\mathbf{p}_{2})(\overset{}{A^{\prime }}_{\overline{f}%
}+\gamma _{5}\overset{}{B^{\prime }}_{\overline{f}})v(\mathbf{p}_{1})\right]
\end{equation}%
where $F_{q}$ is a constant containing CKM factor, $\phi $ is the weak
phase. The amplitude $A_{f}$ and $B_{f}$ are in general complex in the sense
that they incorporate the final state phases $\delta _{p}^{f}$ and $\delta
_{s}^{f}$ and they may also contain weak phases $\phi _{s}$ and $\phi _{p}$.
Note that $A_{f}$ is the parity violating amplitude ($p$-wave) whereas $%
B_{f} $ is parity conserving amplitude ($s$-wave). The $CPT$ invariance
gives the matrix elements for the decay $\bar{B}\rightarrow \bar{N}_{1}N_{2}(%
\bar{f}):$%
\begin{equation}
\bar{M}_{\bar{f}}=F_{q}e^{-i\phi }\left[ \bar{u}(\mathbf{p}%
_{2})(-A_{f}^{\ast }e^{2i\delta _{p}^{f}}+\gamma _{5}B_{f}^{\ast
}e^{2i\delta _{s}^{f}})v(\mathbf{p}_{1})\right]  \label{q2}
\end{equation}%
if the decays are described by a single matrix element $M_{f}$. If $\phi
_{s}=0=\phi _{p}$ then $CPT$ and $CP$ invariance give the same predictions
viz
\begin{equation}
\bar{\Gamma}_{\bar{f}}=\Gamma _{f},\qquad \bar{\alpha}_{\bar{f}}=-\alpha
_{f},\qquad \bar{\beta}_{\bar{f}}=-\beta _{f},\qquad \bar{\gamma}_{\bar{f}%
}=\gamma _{f}  \label{q3}
\end{equation}%
\ In order to test these predictions, consider for example the decay%
\begin{eqnarray}
B_{d}^{0} &\rightarrow &p\overline{\Lambda }_{c}^{-}\rightarrow p\overline{p}%
K^{0} \\
\overline{B}_{d}^{0} &\rightarrow &\overline{p}\Lambda _{c}^{+}\rightarrow
\overline{p}p\overline{K}^{0}
\end{eqnarray}%
By analysing the final states $p\overline{p}K^{0}$, $\overline{p}p\overline{K%
}^{0}$\ one may test $\overline{\alpha }_{\overline{f}}=-\alpha _{f}$\ for
the chamed hyperon (antihyperon) decays.

\section{Conclusion}

\begin{enumerate}
\item Discrete symmetries are not universal both $C$ and $P$ are violated in
weak interaction but respected by electromagnetic and strong interaction.
Violation of $C$ and $P$ are incorporated in the basic structure of weak
interaction by assigning left-handed fermions to the doublet and
right-handed fermions to the singlet representation of electroweak
unification group.

\item Unlike $C$ and $P$ violation, $CP$ violation does not embrace all weak
processes. $CP$\ violation is observed in the semi-leptonic and weak
hadronic decays of mesons.

\item Effective weak interaction Lagrangian in the standard model can
accommodate $CP$\ violation to mismatch between mass eigenstates and CP
eigenstates and/or mismatch between weak eigenstates and mass eigenstates at
quark level. The mixing induced $CP$\ violation involves the mass difference
$\Delta m_{B}$ and $\Delta m_{B_{s}}$.

\item There is no evidence of $CP$\ violation in lepton sector and in
processes involving neutral currents. The effective weak interaction
Lagrangian of the standard model cannot accommodate $CP$ violation in these
sectors. Any experimental observation of $CP$ violation in these sectors
would indicate, physics beyond the standard model.

\item There is no evidence of $CP$ violation in $D^{0}-\overline{D}^{0}$\
complex. $D^{0}$ and $\overline{D}^{0}$\ being bound states of first and
second generation quark and anti-quark; no weak phase in CKM\ matrix is
available to generate $CP$ violation in $D^{0}-\overline{D}^{0}$\ complex.
Any observation of $CP$ violation in these sector would indicate, physics
beyond the standard model.

\item With three generations of quarks, with one phase no extra phase is
available to generate the mismatch between $CP$ and mass eigenstates for $B_{s}^{0}-\bar{B}_{s}^{0}$ complex. The
mixing induced $CP$\ asymmetries for the decays $B_{s}^{0}\rightarrow J/\psi
\phi $ and $B_{s}^{0}\rightarrow K^{+}D_{s}^{(\ast )-}(K^{-}D_{s}^{(\ast
)+}) $%
\begin{eqnarray*}
A_{J/\psi \phi } &=&-\sin 2\beta _{s}\frac{\Delta m_{B_{s}}^{{}}/\Gamma _{s}%
}{1+(\Delta m_{B_{s}}^{{}}/\Gamma _{s})^{2}} \\
&=&0\text{, }\beta _{s}=0\text{ in standard model}
\end{eqnarray*}%
\begin{eqnarray*}
-\left( \frac{S_{+}+S_{-}}{2}\right) _{D_{s}^{(\ast )}K} &\propto &\sin
(2\beta _{s}+\gamma ) \\
&=&\sin \gamma \text{, in standard model}
\end{eqnarray*}%
The experimental determination of these $CP$\ asymmetries in future
experiments when enough data on $B_{s}^{0}$\ decays would be available will
be crucial for any extension of the standard model from three generations of
fermions to four generations of fermions.
\end{enumerate}

Finally for baryon genesis, both $C$ and $CP$ violation are required. How
the $CP$\ violation in meson sector is related to $CP$ violation required
for baryongenesis? There is no answer to this question yet.

\ \

For a review see for instance refs (1-5)

\end{document}